\documentclass{aa}  

\usepackage{multirow}
\usepackage{amsmath}
\usepackage{bm}  
\usepackage{graphicx}
\usepackage{txfonts}
\usepackage[]{hyperref}
\usepackage{CJKutf8}

\newcommand\orc[1]{\href{https://orcid.org/#1}{\includegraphics[width=3mm]{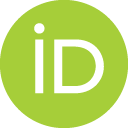}}}

\usepackage{xcolor}  

\newcommand{\ca}[1]{\textcolor{black}{#1}} 
\newcommand{\LG}[1]{\textcolor{black}{#1}}

\newcommand{\connytwo}[1]{\textcolor{black}{#1}} 
\newcommand{\LGtwo}[1]{\textcolor{black}{#1}} 

\newcommand{\CNnames}[1]{{\begin{CJK}{UTF8}{gbsn}~(#1)~\end{CJK}}}

\newcommand{\MISTage}{$102\pm15\,\mathrm{Myr}$}
\newcommand{\jointage}{$132 \pm 8\,\mathrm{Myr}$}
\newcommand{\selfisochroneage}{$129^{+28}_{-23}\,\mathrm{Myr}$}

\begin{document}

   \title{Asteroseismology of the young open cluster NGC 2516}

   \subtitle{II. \LG{Constraining cluster age using gravity-mode pulsators}}

   \author{Gang Li\CNnames{李刚}
          \inst{1}\orc{0000-0001-9313-251X}
          \and
          Joey S. G. Mombarg\inst{2}\orc{0000-0002-9901-3113}
          \and
          Zhao Guo\orc{0000-0002-0951-2171}
          \inst{1}
          \and
     Conny Aerts\inst{1,3,4}\orc{0000-0003-1822-7126}
     }

   \institute{Institute of Astronomy (IvS), Department of Physics and Astronomy, KU Leuven, Celestijnenlaan 200D, 3001 Leuven, Belgium\\
              \email{gang.li@kuleuven.be, conny.aerts@kuleuven.be}
         \and
             Université Paris-Saclay, Université de Paris, Sorbonne Paris Cité, CEA, CNRS, AIM, F-91191 Gif-sur-Yvette, France\\
             \email{Joey.Mombarg@cea.fr}
                          \and
             Department of Astrophysics, IMAPP, Radboud University Nijmegen,
PO Box 9010, 6500 GL Nijmegen, The Netherlands
\and
Max-Planck-Institut für Astronomie, Königstuhl 17, D-69117 
              Heidelberg, Germany
             }

\date{Received ??, ??; Accepted ??, ??}

 
  \abstract
   {Although asteroseismology is regarded as the most powerful tool for probing stellar interiors, seismic modelling remains dependent on global stellar parameters such as temperature and luminosity. Stellar clusters offer direct measurements of these parameters \LGtwo{by fitting a colour–magnitude diagram (CMD)}, making the application of asteroseismology in stellar clusters a valuable approach to advancing the entire field of stellar physics modelling.}
   {We aimed to develop seismic modelling for gravity-mode pulsators in the open cluster NGC\,2516 to determine stellar ages and investigate internal mixing processes.}
   {We computed one-dimensional stellar evolutionary models using the Modules for Experiments in Stellar Astrophysics (\texttt{MESA}), incorporating rotation-induced transport processes. Exponential overshooting in the transition layers between convective and radiative regions was included, as well as rotationally induced mixing in the radiative envelope. Grids of evolutionary models were computed covering isochrone-derived mass ranges. The models were evolved up to 300\,Myr because of the cluster’s young age ($\sim100\,\mathrm{Myr}$).}
   {By fitting the frequencies of identified modes of four gravity-mode member pulsators simultaneously, we measure the seismic age of the cluster NGC\,2516 as \(132 \pm 8 \,\mathrm{Myr}\). This high-precision seismic age estimate deviates by 
   $1\sigma$ from the isochronal age derived from public \texttt{MESA} Isochrones and Stellar Tracks (\texttt{MIST}) isochrones for rotating stars. Our findings show that seismic modelling strongly constrains core overshooting, but because the period spacing patterns are smooth, it provides weak constraints on mixing in the radiative envelope of young gravity-mode pulsators. The two most massive gravity-mode pulsators have \texttt{MIST} masses $\sim\!2.0\,\mathrm{M_\odot}$ while their seismic masses are $1.75\,\mathrm{M_\odot}$. 
We constructed new asteroseismology-calibrated isochrones using input physics identical to that of our seismic model grid. While this resolves the age discrepancy, the mass discrepancy is only partially addressed. The remaining small yet persisting mass discrepancy implies a mismatch between the physics in core to surface environments of 1-D stellar models and the seismic observables probing those areas of fast-rotating stars.}
   {}

   \keywords{Asteroseismology --
                Stars: early-type --
                Stars: interiors --
                Stars: oscillations --
                Stars: rotation --
                open clusters and associations: individual: NGC\,2516
               }

   \maketitle
%


\section{Introduction}

Although one-dimensional (1-D) stellar models have achieved significant success, substantial uncertainties remain in modern stellar evolutionary modelling. One of the major concerns is stellar rotation \citep{Maeder2009}. Rotation plays an important role in modifying a star’s thermal equilibrium and internal physics, especially in fast-rotating early-type stars \citep{von_Zeipel1924, Heger2005}. 
Modelling fast-rotating \ca{stars} should be considered at least a 2-D problem (symmetric about the rotation axis), which might be sophisticated and time-consuming. 
\ca{
Recently, 2-D stellar evolution models have been achieved for rapidly-rotating B-type stars deformed by the centrifugal force by \cite{Mombarg2023, Mombarg2024}, building on the foundation of steady-state models computed with the \texttt{ESTER} code \citep{Espinosa_Lara2011,Rieutord2016}.}


\ca{In contrast to these 2-D evolutionary computations, hydrodynamical simulations covering multiple dynamical timescales have been achieved in two and three dimensions for various masses and evolutionary stages. Their purpose was to simulate angular momentum (AM) transport and chemical mixing by internal gravity waves in intermediate-mass stars \citep[see the series of papers by ][] {Edelmann2019,Horst2020,Varghese2023,Ratnasingam2023,Vanon2023,Herwig2023,Thompson2024,Varghese2024,Rogers2025}. However, such sophisticated simulations cannot be done to cover a star's entire life. A} more time-efficient way to model rotating stars is to acquiesce to some extent 1-D approximations and simplifications \ca{of reality}, such as the shellular approximation
\ca{of rotation and a diffusive approximation of the transport processes \citep{Endal1978,Meynet1997, Heger2000, Heger2005}. Such an approach ignores meridional circulation and advective motions in general \citep{Eggenberger2008, Potter2012}. These simplifications are }
applied in the code  
\ca{
Modules for Experiments in Stellar Astrophysics \citep[\texttt{MESA},][]{Paxton2011ApJS, Paxton2013ApJS, Paxton2015ApJS, Paxton2018ApJS, Paxton2019ApJS, Jermyn2023ApJS}.
Even in the case of more complex rotation-induced advective motions and other induced magnetic phenomena, one can 
adopt 1-D formulae for the transport processes
\citep[e.g.][]{Zahn1992, Meynet1997,Heger2000, 
Spruit2002, Ekstrom2012,Lagarde2012,Georgy2013,Fuller2019,Eggenberger2022}.}

Asteroseismology, the study of oscillation \ca{modes or} waves travelling inside stars \citep{Aerts2010book}, has become a unique and powerful way to study stellar internal structures. The capacity of asteroseismology \ca{to probe} stellar physics \ca{is being exploited} in two ways. First, \ca{observed} stellar oscillation signals carry information about the star's internal environment, such as 
\ca{chemical \citep{Pedersen2018} and}
temperature gradients \citep{Michielsen2021}.
\ca{This allows us to estimate core}
overshooting \citep{Mombarg2021}, near-core rotation \citep{VanReeth2016_TAR, Li2020MNRAS_611}, 
\ca{envelope} mixing profiles \citep{Michielsen2019, Pedersen2021}, and magnetic fields \citep{Li2022Nature, LiGang2023_13_magnetic_RGB, Deheuvels_2023, Hatt2024}. Second, we can constrain and calibrate \ca{some of} the parameterised 
\ca{internal profiles used in}
1-D models using seismic signals. For example, the angular momentum transport has been studied and calibrated using gravity-mode oscillations in A- and F-type main-sequence stars \citep[known as $\gamma$\,Doradus ($\gamma\,$Dor) stars,][]{Kaye1999, Ouazzani2019A&A, Mombarg2023calibrating_AM, Moyano2023, Mombarg2024}.
\ca{The current paper focuses on such intermediate-mass $\gamma\,$Dor pulators.}

Asteroseismic modelling of $\gamma$\,Dor stars 
\ca{is a high-dimensional fitting problem \citep{Aerts2018ApJS} and therefore} relies heavily on \LGtwo{additional (non-seismic)} constraints to lift degeneracies. Spectroscopic observations are necessary, but the parameters derived from spectra (\ca{effective} temperature, luminosity, and surface gravity) provide only loose constraints on stellar evolutionary stages, as their uncertainties are comparable to the \LGtwo{location changes} caused by stellar evolution during the main sequence. As a result, attention turned to multiple systems, with binary stars being the primary focus \citep[e.g.][]{Huber2015, Murphy2025}. Current modelling of eclipsing binaries can achieve uncertainties as low as 0.2\% in masses and radii \citep[e.g.][]{Maxted2020}. Many $\gamma$\,Dor-type components have been found in binary systems \citep[e.g.][]{Schmid2015_10080943_obs,Keen2015_10080943, Li2020_gdor_in_EB, Sekaran2020}, and \ca{asteroseismic binary} modelling work has been achieved \ca{for a handful of systems}, showing the great power of binary-asteroseismology \ca{as a performant} synergistic \ca{approach} \citep[e.g.][]{Schmid2016_10080943_modelling,Guo2017,Guo2019,Zhang2018,Johnston2019,Sekaran2020, Sekaran2021_9850387, Kemp2025}. Beyond the constraints on masses and radii, some studies also highlight the contribution of \ca{the evoluionary stage}, noting that the two pulsating stars in a binary system should share the same age \citep[e.g.][]{Schmid2016_10080943_modelling, Johnston2019}. For example, \cite{Schmid2016_10080943_modelling} required the two 
\ca{$\gamma\,$Dor}
components to have the same age when modelling KIC\,10080943. It is noteworthy that some modelling studies of evolved solar-like oscillators in binary systems have \ca{also} demonstrated age consistency between the two oscillating components \citep[e.g.][]{Metcalfe2015, White2017_binary, LiYaGuang2018_solarlike_binary}. The equal-age assumption and the mass and radius constraints provided by binary systems greatly facilitates the asteroseismic models.

Not only binary systems, but also stellar clusters — as multi-star systems — provide additional constraints for asteroseismic modelling. Stellar parameters of member stars in clusters can be measured using the colour–magnitude diagram (CMD), as they share \LGtwo{nearly} the same distance, age, and \ca{initial} metallicity. Several studies have identified pulsating stars in clusters. \ca{High-precision asteroseismology has been developed for} the four open clusters in the Kepler field \citep[NGC\,6866, 6819, 6791, 6811, see e.g.:][]{Corsaro_2012, Balona2013}, as well as for M67 and M4 in the K2 field \citep{Miglio2016_M4, Stello2016_M67}. \ca{More recently, the young open clusters} UBC-1, Pleiades, and NGC\,2516 in the TESS continuous viewing zone 
\ca{were scrutinised observationally}
\citep{Fritzewski2024, Bedding2023ApJ, LiGang_2024_NGC2516}, \LGtwo{as well as some stellar associations \citep{Kerr2022SPYGLASSII,Kerr2022SPYGLASSIII} and star-forming complexes \citep{Murphy2024Cep_Her_complex}.} Asteroseismic modelling efforts in clusters have mainly focused on solar-like oscillators, including M67 \citep{Reyes2025MNRAS}, M4 \citep{Tailo_2022_M4}, NGC\,6866 by \cite{Brogaard2023}, and more \citep[see:][]{Tayar2025}. These studies have revealed some discrepancies in age and mass estimates compared to isochrone fitting results \ca{relying on precomputed stellar models to fit the stars in the CMD without taking into account asteroseismic input}. Modelling of young open clusters using pulsating main-sequence stars remains scarce.

In this paper, we present the first \ca{joint} asteroseismic modelling of main-sequence gravity-mode pulsators in a young open cluster. The structure of the paper is as follows. Section~\ref{sec:sample_obs_constraints} introduces previous pilot studies of pulsating stars in the open cluster NGC\,2516, as well as the observational constraints on the member stars derived from the best-fitting isochrone \ca{in the CMD}. 
Section~\ref{sec:model_construction} describes the input physics adopted in our asteroseismic modelling, the construction of \ca{our novel} model grid, and the search for the best-fitting models. We apply two approaches. First, \ca{we} allow the stars to have independent ages. \ca{In second instance, we} enforce a common age for all \ca{considered pulsators}. 
In Section~\ref{sec:results}, we present our results from both approaches and report discrepancies in age and mass from different methods. 
To address these discrepancies, we construct \ca{novel} asteroseismology-calibrated isochrones in Section~\ref{sec:own_isochrone}. \ca{We} demonstrate that the 
\ca{cluster's asteroseismic isochrone solves the} age discrepancy, while \ca{it does not resolve a modest remaining mass discrepancy}. 
Conclusions are given in Section~\ref{sec:conclusions}.

\section{The sample and observational constraints}\label{sec:sample_obs_constraints}

\cite{LiGang_2024_NGC2516} studied the photometric variability of member stars in the open cluster NGC\,2516. In total, 24 stars were found to exhibit gravity-mode (g-mode) pulsations. These stars include $\gamma$\,Dor stars at the low-temperature end \citep[with $T_\mathrm{eff}\sim 7000\,\mathrm{K}$][]{Balona1994,Kaye1999,Dupret2005}, Slowly Pulsating B-type (SPB) stars \citep{Waelkens1991, DeCat2002A&A}, and some g-mode pulsators that appear between the blue edge of theoretical instability strips (IS) of $\gamma$\,Dor stars and the red edge of the IS of SPB stars \citep{Dupret2005A&A, Szewczuk2017}, which have been noticed both in pre-Gaia ground-based photometry and by the Gaia mission \citep{Mowlavi2013, DeRidder2023A&A, Mombarg2024AA_14000Gaia}. Considering the similarities in observational properties and internal physical mechanisms among these variable stars, we collectively refer to them as `g-mode pulsators', without further subclassification based on spectral type.

Among the 24 g-mode pulsators in NGC\,2516, 11 show clear period spacing patterns. These patterns refer to the period differences ($\Delta P = P_{n+1,l,m}-P_{n,l,m}$) between consecutive modes ($P_{n,l,m}$) with the same angular degree $l$ and azimuthal order $m$, but with radial orders $n$ differing by $\pm1$. \LGtwo{In fact, if the period-spacing pattern is interrupted, we can also fit the periods themselves instead of the period spacings
\connytwo{as discussed in great detail by \citet{Michielsen2021}}. 
The presence of such clear period spacing patterns enables us to model the g-mode pulsators in more detail. }

We selected our sample for seismic modelling based on the results of \cite{LiGang_2024_NGC2516}. Although 11 g-mode stars exhibit clear period spacing patterns, not all of them are \ca{equally} suitable for \ca{forward} modelling, as some deviate from the best-fitting isochrone. These deviations may be caused by binarity, which introduces systematic deviations in the determination of stellar parameters such as effective temperature, luminosity, and mass. Therefore, we selected only six stars that lie exactly on the best-fitting isochrone, as listed in Table~\ref{tab:observational_constraints}.

\cite{LiGang_2024_NGC2516} used the \texttt{MIST} isochrones \citep[\texttt{MESA} Isochrones and Stellar Tracks,][]{Dotter2016ApJS_MIST,Choi2016ApJ_MIST} to fit the colour–magnitude diagram (CMD) of NGC\,2516 and derived age and extinction. \LGtwo{They tested different rotation rates and found that the slower-rotating isochrones provided the best fit, as they resulted in the smallest residuals.} Following their work, \LGtwo{we derived the effective temperatures, luminosities, and masses of the six selected g-mode pulsators from their best-fitting MIST isochrone. \connytwo{The values for these three quantities of the six pulsators then} form the basis for constructing the model grid and serve as observational constraints in our subsequent modelling.} The results are presented in Table~\ref{tab:observational_constraints}. The best values of temperatures, luminosities, and masses are \ca{taken as those of} the nearest point in the best-fitting isochrone, while the uncertainties were estimated based on the scatter in the CMD. Although it is commonly assumed that member stars in a cluster share the same distance, extinction, and age, the main sequence in the CMD is not an infinitely narrow line. Instead, it exhibits scatter due to several external factors, including photometric errors, differential extinction, binarity, variations in internal stellar physics (primarily the extent of overshooting and rotation-induced envelope mixing), and differences in stellar inclination, which affect gravity darkening. We estimated that the scatter in the Gaia G-band magnitude (the y-axis of the CMD) is 0.146\,mag, and used this to propagate uncertainties to the effective temperatures, luminosities, and masses.

\begin{figure}
    \centering
    \includegraphics[width=0.9\linewidth]{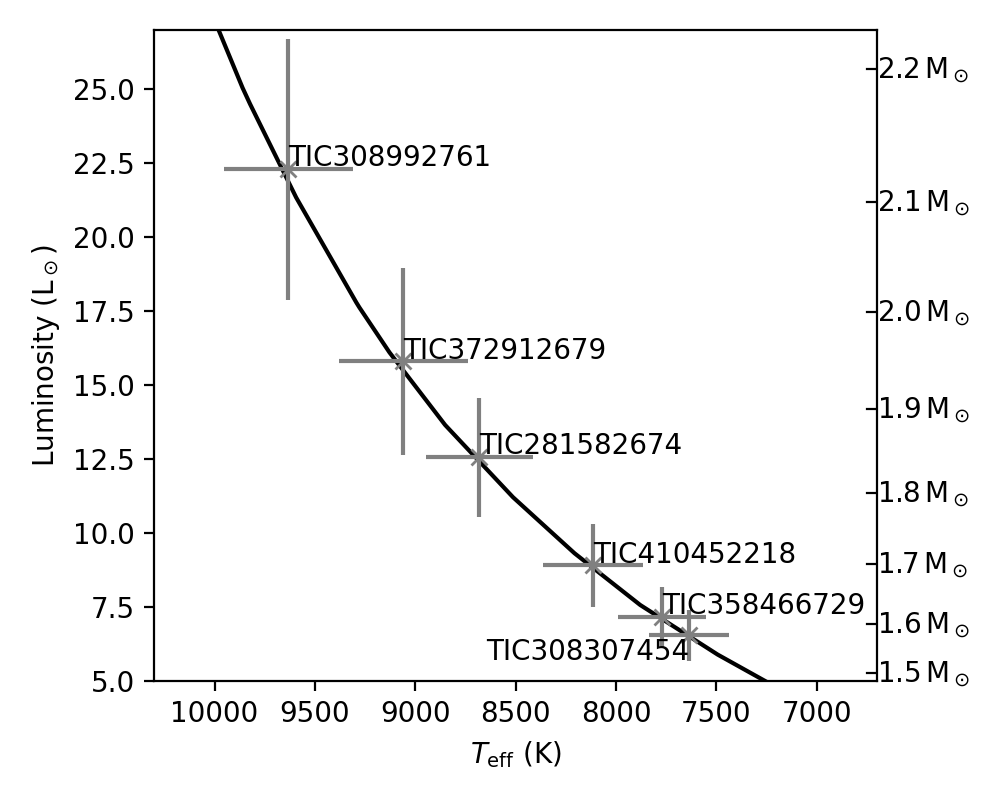}
    \caption{Isochrone-derived luminosities and effective temperatures of the selected $\gamma$\,Dor stars in NGC\,2516. Error bars show the 1$\sigma$ level of the uncertainties. The solid curve is the best-fitting \texttt{MIST} isochrone. }
    \label{fig:observational_constraints}
\end{figure}

\LGtwo{In Fig.~\ref{fig:observational_constraints}, we plot the six selected g-mode pulsators in the Hertzsprung–Russell (HR) diagram, with error bars indicating the uncertainties in effective temperature and luminosity. Intuitively, one might expect the brighter stars to have lower uncertainties, since their photometric errors are smaller. However, as explained above, the uncertainties in $T_\mathrm{eff}$ and $L$ were derived from the scatter in the CMD. Because the best-fitting isochrone becomes steeper on the high-temperature side, the resulting uncertainties in luminosity -- and consequently in mass -- are larger for the hotter pulsators than for the cooler ones.}

\begin{table}[]
    \centering
        \caption{Isochrone-derived observational constraints \ca{for the} effective temperatures, luminosities, and masses of the six $\gamma$\,Dor stars in NGC\,2516.}
\small
\begin{tabular}{cccc}
\hline
TIC	& $T_\mathrm{eff}$ (K) & $L$ ($\mathrm{L_\odot}$)  & $M$ ($\mathrm{M_\odot}$)\\
\hline
308307454 & $7626 \pm 220$ & \phantom{0}$6.5 \pm 1.0$  & $1.57 \pm 0.05$\\
358466729 & $7760 \pm 240$ & \phantom{0}$7.1 \pm 1.1$ & $1.61 \pm0.06$\\
410452218 & $8220 \pm 260$ & \phantom{0}$9.5 \pm 1.5$ & $1.72 \pm 0.06$ \\
281582674 & $8700 \pm 300$ & $12.7 \pm 2.4$ & $1.84 \pm 0.08$ \\
372912679 & $9090 \pm 290$ & $16.0 \pm 2.7$ & $1.95 \pm 0.08$ \\
308992761 & $9699 \pm 300$ & $22 \pm 4$ & $2.11 \pm 0.11$ \\
\hline
    \end{tabular}

    \label{tab:observational_constraints}
\end{table}

\section{\ca{Theoretical model predictions}}\label{sec:model_construction}

We used  \LG{\texttt{MESA} version r24.03.1} to compute rotating 1-D stellar structures for the selected g-mode pulsators. At each time step, the resulting stellar structures were passed to the stellar oscillation code \texttt{GYRE} \citep{Townsend2013GYRE, Townsend2018GYRE, Sun2023_GYRE_TIDES} to calculate the oscillation frequencies of g modes under the traditional approximation of rotation \citep[TAR, see:][]{Bildsten1996TAR, Lee1997, Townsend2003, Mathis2013LNP,Saio2018} assuming rigid rotation, which is a good approximation for $\gamma\,$Dor pulsators as found by \citet{Li2020MNRAS_611}. We assume the stars to be non-magnetic in the first instance because the effects of the Lorentz force are expected to be much smaller than those of the Coriolis force \citep{Aerts2021,Dhouib2022, Lignieres2024} except for very strong central fields not yet detected so far \citep{Rui2024}. \connytwo{This might be the result of a magnetic suppression effect as found in slow rotators \citep{Fuller2015Sci, Stello2016Nature, Murphy2020magnetic_damping}. Evidence for such suppression has not yet been found for fast rotators. }

We adopted the same \texttt{MESA} input physics and inlist as used by \cite{Mombarg2024AA_14000Gaia}\footnote{\url{https://zenodo.org/records/13759780}} and the same inlist for \texttt{GYRE} as in \cite{Mombarg2023calibrating_AM}\footnote{\url{https://zenodo.org/records/7737918}}. We refer the readers to these two papers for details. One important note is that we used a higher structural resolution by setting \texttt{mesh\_delta\_coeff  = 0.3}. In addition, we did not set a minimum diffusion (\texttt{set\_min\_D\_mix = .false.}). A summary of the input physics, free parameters, and grid construction adopted in our work is provided below.

\subsection{Input physics of \texttt{MESA} models}\label{susec:input_physics}

Quite often, the rotational effects caused by the Coriolis and centrifugal forces are only treated at the level of pulsation computations applied to a non-rotating equilibrium model \citep{Aerts2021RvMP}. This is justified in the case of slow to modest rotation, because stellar models including rotational effects remain uncalibrated by observations \citep{AT2024}. However, in this work, we have to include a description of rotation in the 1-D stellar models because the selected six g-mode pulsators are rotating rapidly. Based on the observed period spacing patterns, \cite{LiGang_2024_NGC2516} reported their rotation rates to be about 50\% of the Keplerian critical rotation rates. In \texttt{MESA} \ca{versions} as introduced since \cite{Paxton2013ApJS} and \ca{further advanced by} \cite{Paxton2019ApJS}, the shellular approximation is adopted. \ca{This approximation} assumes that the angular velocity is constant over isobars. The coordinate and structure equations are then refined to impose that a sphere with this refined radius has the same volume as the oblate spheroid in the Roche model, to account for the effects of the centrifugal force \citep[see also][]{Endal1976,Meynet1997,Rieutord2016,Paxton2019ApJS}. 

We set the initial fraction of Keplerian critical rotation at the zero-age main sequence (ZAMS), $v_\mathrm{i}/v_\mathrm{i, crit}$, assuming a uniform rotation profile at the ZAMS. Since the stars in NGC\,2516 are young ($\sim 100\,\mathrm{Myr}$), \ca{we assume that their} rotational deceleration is marginal \ca{and thus} adopted the current seismic rotation rates reported by \cite{LiGang_2024_NGC2516} as their initial rotation rates. For TIC\,308307454, we set $v_\mathrm{i}/v_\mathrm{i, crit} = 0.167$, and for TIC\,358466729, $v_\mathrm{i}/v_\mathrm{i, crit} = 0.333$. For the remaining selected stars, which have similar seismic rotation rates, we adopted a common value of $v_\mathrm{i}/v_\mathrm{i, crit} = 0.5$.

\ca{In general, angular momentum transport involves both advective and diffusive components in the transport equation \citep[e.g.,][for a review of the processes]{Aerts2019ARA&A}. Here, the}
angular momentum transport is treated in a fully diffusive way 
\ca{since we use the public \texttt{MESA} code.} We did not adopt a radius-dependent viscosity derived from the built-in rotation- and magnetism-induced mechanisms in \texttt{MESA} \citep{Heger2000, Spruit2002, Heger2005}, as such prescriptions lead to non-smooth viscosity profiles, which are undesirable for asteroseismic modelling. Instead, a constant viscosity of $10^7\,\mathrm{cm^2\,s^{-1}}$ was used, which was calibrated using the slowly rotating $\gamma$\,Dor stars near the terminal-age main sequence (TAMS) \citep{Mombarg2023calibrating_AM}. We also accounted for gravity darkening effects at an average inclination \citep{Espinosa_Lara2011,Rieutord2016,Paxton2019ApJS}, although their influence on effective temperature and luminosity is minor and is smaller than the observational uncertainties \ca{shown in Fig.\,\ref{fig:observational_constraints}.}

\connytwo{Here, we emphasise the purpose of using rotating 1D models with rotation measured by asteroseismology: we expect that the rotating models can correct their internal structure (e.g., temperature and density) by taking the centrifugal force into account. Such a correction can reproduce the evolutionary rates and tracks of real stars as closely as possible. 
We simply used the present-day near-core rotation rates as initial rotation rates. Such a simplification, on the one hand, reduces one free parameter, and on the other hand, allows us to mimic as much as possible the impact of rotation on stellar structure and evolutionary rates. 
Our main aim for the forward modelling of the period spacings is to obtain precise age estimates. As such, we take the present-day near-core rotation frequency of the target stars as a fixed parameter rather than using the predicted rotation profile from the MESA models, as this would introduce additional uncertainties on both the initial rotation profile and the efficiency of AM transport. Considering the present-day rotation rate as a fixed parameter implies that we use a uniform rotation profile for the calculations of the oscillation periods. Indeed, the near-core-to-surface rotation ratios in $\gamma$\,Dor stars are close to 1 \citep{Van_Reeth2018, Li2020MNRAS_611}.}

We described core convection using the mixing-length theory with a solar-calibrated value of $\alpha_\mathrm{MLT} = 1.8$ \citep{Cox1968, Choi2016ApJ_MIST}. \LGtwo{We note that the solar-calibrated value may not be universal, as different calibration methods for the Sun \citep[e.g.][]{Joyce2018_metal_poor} and for non-solar stars \citep[e.g.][]{Metcalfe2012, Joyce2018_alpha_Centauri} show systematic differences. The value of $\alpha_\mathrm{MLT}$ may also vary with stellar mass, metallicity \citep{Bonaca2012, Tayar2017, Viani2018}, and the choices of other input physics \citep[e.g.][]{Cinquegrana2022}. For stars with convective cores (i.e., the g-mode pulsators in this work), the choice of $\alpha_\mathrm{MLT}$ does not have a strong impact, as core convection is sufficiently efficient \citep{Kippenhahn1994book}. However, for F-type stars with shallow convective envelopes, the choice of $\alpha_\mathrm{MLT}$ may still play a role in asteroseismology \citep{Aerts2018ApJS}. We refer the reader to the review by \cite{Joyce2023} for more information and discuss details in Appendix\,\ref{app_sec:MLT}.}

We adopted the exponentially decaying overshoot $f_\mathrm{CBM}$ in unit of local pressure scale height to describe the convective-boundary mixing \citep[CBM,][]{Herwig2000}. In the radiative envelope, the chemical diffusion coefficient $D$ was computed from the rotational mixing based on by \cite{Zahn1992} and \cite{Chaboyer1992},
\begin{equation}
    D_\mathrm{rot}\left(r > r_\mathrm{cc}\right) = \eta K \left(\frac{r}{N} \frac{\mathrm{d}\Omega}{\mathrm{d} r}\right)^2, \label{eq:rotation_diffusion}
\end{equation}
where $K$ is the thermal diffusivity, $N$ is the Brunt--V\"ais\"al\"a frequency, $r_\mathrm{cc}$ is the radius of the convective core, $\Omega$ is the rotation profile, and $\eta$ is a free parameter that modifies the \ca{efficiency of the diffusive transport.} 

Finally, we adopted a solar metallicity of $\mathrm{Z}=0.014$ \citep{Asplund2009}, consistent with the spectroscopic observations of NGC\,2516 by \citep{LiGang_2024_NGC2516}. The present-day cosmic abundances, based on a chemically homogeneous sample of 29 early B-type stars in the solar neighbourhood, were used in our modelling \cite{Nieva2012}. An initial helium abundance $Y = 0.2612$ was set, following the chemical enrichment rate derived by \citep{Verma2019}, which is $Y=0.244+1.226Z$.



\subsection{Model grids for each star}

\begin{table}[]
\tabcolsep=3pt
    \centering
        \caption{Ranges and steps of the free parameters \ca{in the model grid}. }
    \small
    \begin{tabular}{l|l|l}
    \hline
                             & TIC\,308307454 and & The other stars \\
                             & TIC\,358466729 & \\
    \hline
    $M$ range, step  & $\pm2\sigma$, $0.01\,\mathrm{M_\odot}$ & $\pm2\sigma$, $0.02\,\mathrm{M_\odot}$ \\
    \hline
    $\log \eta$ min, max, step    & 0, 7, 1 & 0, 5, 1\\
    \hline
    $f_\mathrm{CBM}$ min, max, step  & \multicolumn{2}{c}{0.005, 0.035, 0.0025} \\
    \hline
    Age $t$ & \multicolumn{2}{c}{up to 300\,Myr}\\
    \hline
    Initial rotation $v_\mathrm{i}/v_\mathrm{i,crit}$ & \multicolumn{2}{c}{Seismic values by \cite{LiGang_2024_NGC2516}}\\
    \hline
    \end{tabular}
    \label{tab:grid_settings}
    \tablefoot{The medians of the mass grids are the observed masses given in Table~\ref{tab:observational_constraints} for each star. The unit of $f_\mathrm{CBM}$ is the local pressure scale height. }
\end{table}

The free parameters for each star are: initial mass $M$, overshoot $f_\mathrm{CBM}$, age $t$, and the mixing parameter $\eta$. 
Table~\ref{tab:grid_settings} summarises the ranges and step sizes of the free parameters adopted in our stellar modelling. Two \ca{sub-grids} are considered: (1) the two low-mass stars TIC\,308307454 and TIC\,358466729, and (2) all other high-mass stars in the sample.  The main differences between these two stem from mass and envelope mixing. For the stellar mass $M$, we varied it within $\pm2\sigma$ around the observed value, using a step size of $0.01\,\mathrm{M_\odot}$ for the two low-mass stars, and $0.02\,\mathrm{M_\odot}$ for the rest. The reason is that the two low-mass stars (TIC\,308307454 and 358466729) have smaller mass uncertainties, so a smaller mass step was used to pursue a higher grid resolution. 

The mixing parameter $\log\eta$ was sampled from 1 to 7 with a step size of 1 for the two low-mass stars, and from 1 to 5 for the others. In the process of determining the range of $\eta$, we gradually increased its upper limit until the stellar evolution exhibited quasi-chemically homogeneous evolution, \ca{which is characterised by a rising temperature during the main-sequence phase}
\LG{because the mixing timescale is shorter than the nuclear
timescale \citep[e.g.][]{Maeder1987, Langer1992, Martins2013quasichemically}.} We found that stars with higher masses tend to exhibit this behaviour at lower values of $\eta$. Therefore, we used $\log\eta$ up to 7 for the two low-mass stars, while limiting it to 5 for the remaining stars.

The range of exponential overshooting parameter $f_{\rm CBM}$ was set from 0.005 to 0.035 with step of 0.0025, covering the previous seismic results for A- to F-type stars \citep{Mombarg2021}. For age $t$, we allowed models to evolve up to 300\,Myr, which is roughly two to three times the ages determined from isochrone fittings (e.g., \MISTage~by \citealt{LiGang_2024_NGC2516};, $240\,\mathrm{Myr}$ by \citealt{Cantat-Gaudin2020}, and $125^{+130}_{-50}\,\mathrm{Myr}$ by \citealt{Hunt2024}). We performed the stellar evolution grid computations on the supercomputers of the \ca{Flemish}  VSC (Vlaams Supercomputer Centrum), using 100 threads with four cores allocated per thread.

\subsection{Best-fitting model search} \label{subsec:best_fitting_model_search}
We {output an internal profile of the stellar structure computed by \texttt{MESA} every time the central hydrogen mass fraction $X_\mathrm{c}$ decreases by 0.01, starting from $X_\mathrm{c} = 0.72$. At each 
$X_\mathrm{c}$, \texttt{GYRE} was used to calculate the g-mode periods \LGtwo{using the rotation rates measured by the observed period spacing patterns by} \citep{LiGang_2024_NGC2516}, \LGtwo{instead of the MESA-calculated theoretical values.} \LGtwo{We did not aim for the rotation rates computed by MESA to match the observations, as explained Section~\ref{susec:input_physics}. Our objective was to correct the temperature, luminosity, and evolution including rotational effects in the most sensible way, given asteroseismic measurements.}

We \ca{used} two methods to search for the best-fitting models \ca{relying on}: (1) a traditional $\chi^2$ \ca{merit function, in our application minimising the difference between the predicted and measured mode periods or mode period spacings}, and (2) the Mahalanobis distance approach \ca{introduced in forward asteroseismic modelling by \citet{Aerts2018ApJS} to take into account theoretical uncertainties in the mode period predictions due to shortcomings in the input physics of 1-D stellar models}. 
The $\chi^2$-based method \ca{often} assumes that the data are Gaussian distributed and that each \ca{of the measured periods or period spacings used in the  merit function is independent from all others}. In contrast, the Mahalanobis distance accounts for correlations \ca{among the measured mode periods or period spacings by incorporating their covariances. Moreover, this merit function also includes the covariances among the theoretically predicted mode periods stemming from the correlations between the free parameters of the stellar models in the grid. As an example, the age of a stellar model in the grid is strongly dependent on the model's core overshooting value, as the latter determines how much fuel can take part in the nuclear burning active in the convective core as the model evolves. Graphical representations of the correlation structure within grids of models used for g-mode asteroseismology are available in 
\citet{Moravveji2015,Michielsen2019,Michielsen2021}.} 

\ca{While the principle of taking into account theoretical uncertainties in the merit function is uncontested in the era of space asteroseismology \citep{Gruberbauer2012}, its practical application is challenging as we do not know to what extent the physical ingredients captured by free parameters in the models deviate from reality. Following \citet{Aerts2018ApJS}, this theoretical uncertainty, which is much larger than the observational uncertainties, can be estimated from the grid of models itself, provided that the grid is a reasonable representation of the star's global and seismic properties covering the uncertainty ranges of the observables.  Since we seldom know the age or evolutionary stage of a g-mode pulsator in the field of the galaxy with good precision, \citet{Pedersen2021} and \citet{Michielsen2021,Michielsen2023} used the correlation structure within their grids of evolutionary models from the birth to the exhaustion of hydrogen in the core in their g-mode asteroseismology. In our application, however,
the stellar models were only evolved up to 300\,Myr as the cluster's age is roughly known. This prevents us from properly \LGtwo{assessing} the correlation structure within the grid of models.}
As a result, the Mahalanobis distance yields too loose constraints on the 
\ca{minimisation process evaluating the model predictions.} 
\ca{We therefore ignore the off-diagonal covariance elements and retain only the diagonal variance terms for both the observations and theoretical predictions.} In that case the Mahalanobis distance reduces to a $\chi^2$-type statistic, which can be written as
\begin{equation}
    \chi^2_j = \left( \boldsymbol{Y}_j^\mathrm{theo} - \boldsymbol{Y}^\mathrm{obs} \right)^\mathrm{T} \left(\boldsymbol{\mathrm{V}} + \boldsymbol{\Sigma} \right)^{-1} \left( \boldsymbol{Y}_j^\mathrm{theo} - \boldsymbol{Y}^\mathrm{obs} \right).
\end{equation}
\ca{Here, $\chi^2_j$ is the chi square of the $j^\mathrm{th}$ model and $\boldsymbol{Y}_j^\mathrm{theo}$ is a vector containing the corresponding theoretical grid values of the observables $\boldsymbol{Y}^\mathrm{obs}$, which can be either mode periods or mode period spacings. The diagonal matrix $\boldsymbol{\mathrm{V}}$ contains the theoretical uncertainties assessed from the grid's mode prediction variances}.

\ca{Following \citet{Michielsen2021}, the} likelihood function is defined as 
\begin{equation}
    \mathcal{L}\left(D |\theta \right) = \exp\left(-\frac{1}{2}\ln\left(|\mathrm{V+\Sigma}|\right) + k \ln \left(2\pi \right) + \chi_{j}^2 \right),\label{eq:likelihood}
\end{equation}
where $\mathcal{L}\left(D |\theta \right)$ denotes the likelihood of the data $D$ given the parameters $\theta$, $|\mathrm{V+\Sigma}|$ is the determinant of the sum of the matrix $\boldsymbol{\mathrm{V}}$ and $\boldsymbol{\mathrm{\Sigma}}$, and $k=4$ is the number of free parameters, namely the mass $M$, the mixing parameter $\eta$, the overshoot $f_\mathrm{CBM}$, and the age $t$. 
The observational uncertainties, represented by the elements of $\boldsymbol{\Sigma}$, are typically much smaller than the model uncertainties, which are written in the diagonal elements of the matrix $\mathbf{V}$. Therefore, it is necessary to develop an appropriate method for estimating the 
\ca{uncertainties for the mode periods predicted theoretically from the grid, of relevance for each of the pulsators. Our approach} 
is to estimate the uncertainty arising from variations in stellar mass and age. Specifically, we calculated the standard deviations of the selected theoretical values of the periods or period spacings within a narrow parameter range: a mass variation $\Delta M$ of $0.02\,\mathrm{M_\odot}$ and an age variation $\Delta t$ of $10\,\mathrm{Myr}$. The range in mass was determined by the step size of the input grid, while the range in age was derived from the uncertainty given by the isochrone fitting. 
We found that within these ranges, the model variations provide a reasonable estimate of the theoretical uncertainty, while also offering sufficient constraints to identify the best-fitting models. In this way, the matrix $\mathbf{V}$ serves both to constrain the model selection and to quantify the model uncertainty. 
We also experimented with including variations in the overshooting parameter, namely by adding another dimension with $\Delta f_\mathrm{CBM} \leq 0.0025$, but this resulted in overly loose constraints that prevented the identification of best-fitting models. 
In addition, we tested the impact of the mixing parameter $\eta$ and found that it has little influence on the modelled periods. 

We calculated the likelihood function at each evolutionary step using Eq.~\ref{eq:likelihood}, and then ran a Markov Chain Monte Carlo (MCMC) simulation using the Python package \texttt{emcee} \citep[][]{2013PASP..125..306F} to search for the best-fitting models. Flat priors for $T_\mathrm{eff}$ and $L$ were adopted within the $2\sigma$ ranges of the observational constraints listed in Table~\ref{tab:observational_constraints}, while a flat prior for age $t$ was set between 50 and 300\,Myr. 

For the observables \(\boldsymbol{Y}^\mathrm{obs}\) and the corresponding theoretical values \(\boldsymbol{Y}_j^\mathrm{theo}\), we used g-mode periods for most of the stars. However, we found that the period spacing pattern of TIC\,281582674 exhibits a significant dip, and using periods as observational constraints could not efficiently reproduce the dip. Therefore, we used period spacings as observables for TIC\,281582674 instead.

\subsection{Individual or joint fitting \ca{of modes}}\label{subsec:individual_or_joint_fitting}

We applied two fitting approaches to search for the best-fitting models. In the first approach, we fitted the stars individually, allowing them to have different ages despite being located within the same cluster. In the second approach, we \ca{demanded} the age to be the same for all targets and constructed the joint likelihood function by summing the logarithms of the individual likelihoods of each star, namely,
\begin{equation}
    \log \mathcal{L}\left(D' \mid \theta', t \right) = \sum_i \log \mathcal{L}\left(D_i \mid \theta_i, t \right),\label{eq:joint_likelihood}
\end{equation}
where \(D'\) represents the combined observables of all stars, \(\theta'\) denotes the collection of individual stellar parameters \(M\), \(f_\mathrm{CBM}\), and \(\eta\) for all stars, $\theta_i$ denotes the free parameters for each star except for the age $t$, and \(t\) is the common age. Approach~1 allows additional freedom in age, which generally leads to a better reproduction of the observed period spacing patterns. In contrast, Approach~2 is more consistent with the physics of open clusters, where stars are generally assumed to have formed simultaneously.

\section{Results \ca{of the asteroseismic modelling}}\label{sec:results}

\begin{table*}[]
    \centering
        \caption{Best-fitting parameters for the successfully fitted stars. }
    \small
    \begin{tabular}{c|cccc|cccc}
    \hline
             & \multicolumn{4}{c|}{Approach 1} & \multicolumn{4}{c}{Approach 2} \\
    TIC      & $M$ ($\mathrm{M_\odot}$) & $f_\mathrm{CBM}$ & $\log \eta$ & $t$ (Myr) & $M$ ($\mathrm{M_\odot}$) & $f_\mathrm{CBM}$ & $\log \eta$ & $t$ (Myr) \\
    \hline
TIC308307454 & 1.537(8) & 0.0051(25) & 1.8(1.4) & 152(18) & 1.5306(23) & 0.0050(25) & 2.5(1.5) & \multirow{4}{*}{132(8)}\\ 
TIC358466729 & 1.61(3) & 0.013(5) & 2.5(1.8) & 180(70) & 1.612(28) & 0.011(4) & 2.4(1.7) & \\ 
TIC410452218 & 1.746(23) & 0.013(4) & 2.0(1.5) & 150(70) & 1.751(23) & 0.012(4) & 2.4(1.5) & \\ 
TIC281582674 & 1.752(18) & 0.0296(25) & 2.1(1.6) & 111(15) & 1.770(25) & 0.0297(25) & 2.2(1.6) & \\ 
\hline
    \end{tabular}
    \label{tab:best_fitting_results}
    \tablefoot{Results from both Approach 1 (individual fitting) and Approach 2 (joint fitting assuming a uniform age) are provided. The numbers in parentheses indicate the uncertainties in the last digit(s) of the corresponding values.}
\end{table*}

We list the best-fitting parameters and their uncertainties in Table~\ref{tab:best_fitting_results}, derived from both Approach~1 and Approach~2. Since the parameters are \ca{selected from 
discrete grid points}, the traditional method of estimating the uncertainties and median values using the 16th, 50th, and 84th percentiles may be unsuitable. Therefore, we fitted a Gaussian distribution to each of the marginalised one-dimensional posterior distributions to derive their uncertainties. 
In cases where the standard deviation of the fitted Gaussian is smaller than the grid step size, we adopt the grid step as the uncertainty. This situation often arises for the overshooting parameter $f_\mathrm{CBM}$, indicating that the modelling is overly sensitive to its value. We conclude that the current grid step of 0.0025 is still too large to capture the sensitivity adequately.

We present two representative stars, TIC\,308307454 and TIC\,281582674, as case studies in Sections~\ref{subsec:308307454} and \ref{subsec:281582674}, respectively. The results on the seismic age of the open cluster, the constraints on $f_\mathrm{CBM}$, and the influence of the mixing parameter $\eta$ are discussed after that. In Table~\ref{tab:best_fitting_results}, we include only four stars for which satisfactory fits were obtained. The two high-mass stars, TIC\,372912679 and TIC\,308992761, could not be fitted successfully, and we discuss the possible reasons in Section~\ref{subsec:mass_discrepancy}.

\subsection{TIC\,308307454}\label{subsec:308307454}
\begin{figure}
    \centering
    \includegraphics[width=0.9\linewidth]{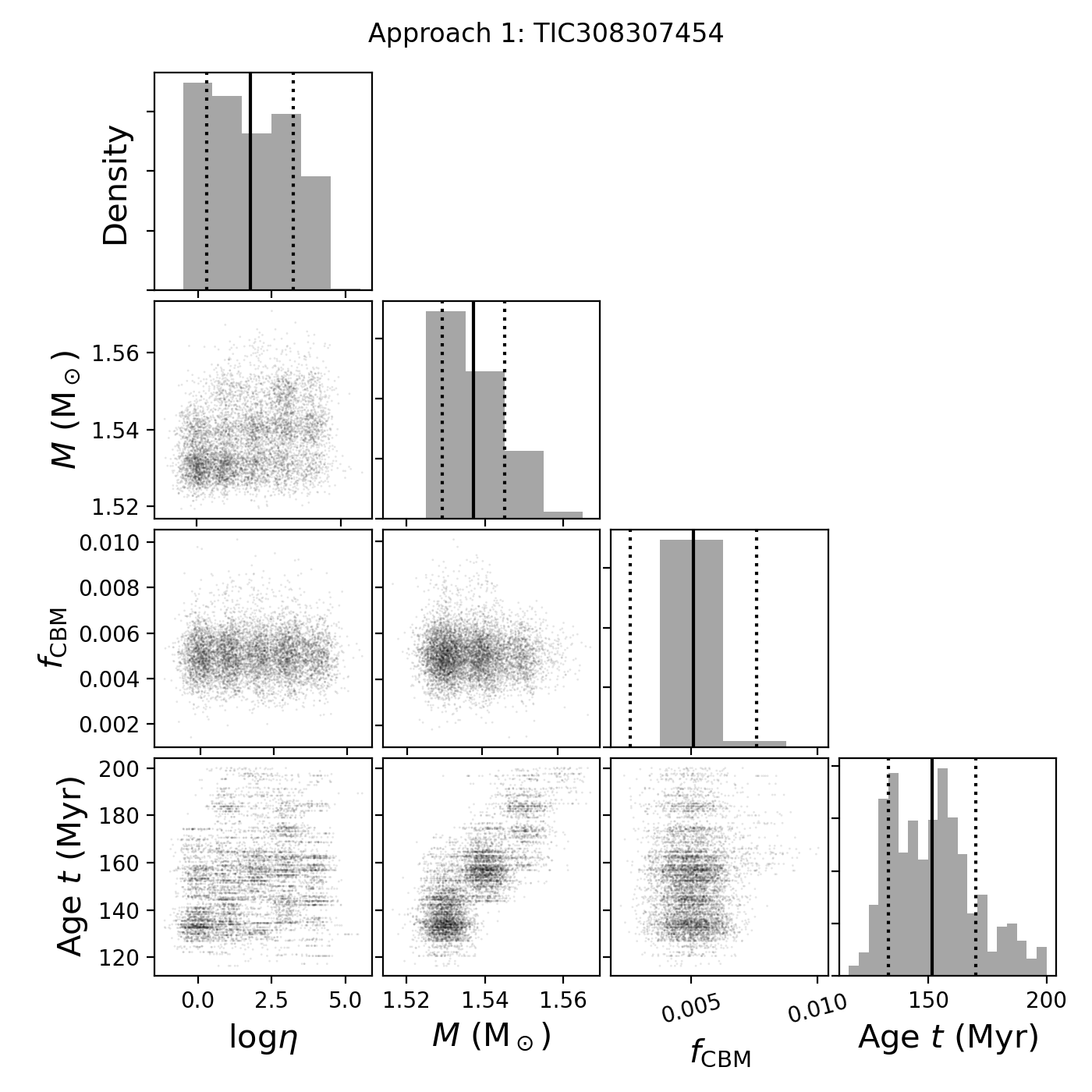}
    \caption{Posterior distributions of the model parameters for TIC\,308307454 obtained using Approach 1. The diagonal panels show the marginalised one-dimensional distributions for each parameter, where the solid and dashed vertical lines indicate the median and the $\pm1\sigma$ range, respectively. The off-diagonal panels display the two-dimensional joint posterior distributions. We added random noise to the discrete samples to enhance their visual clarity. }
    \label{fig:corner_TIC308307454}
\end{figure}

\begin{figure}
    \centering
    \includegraphics[width=1\linewidth]{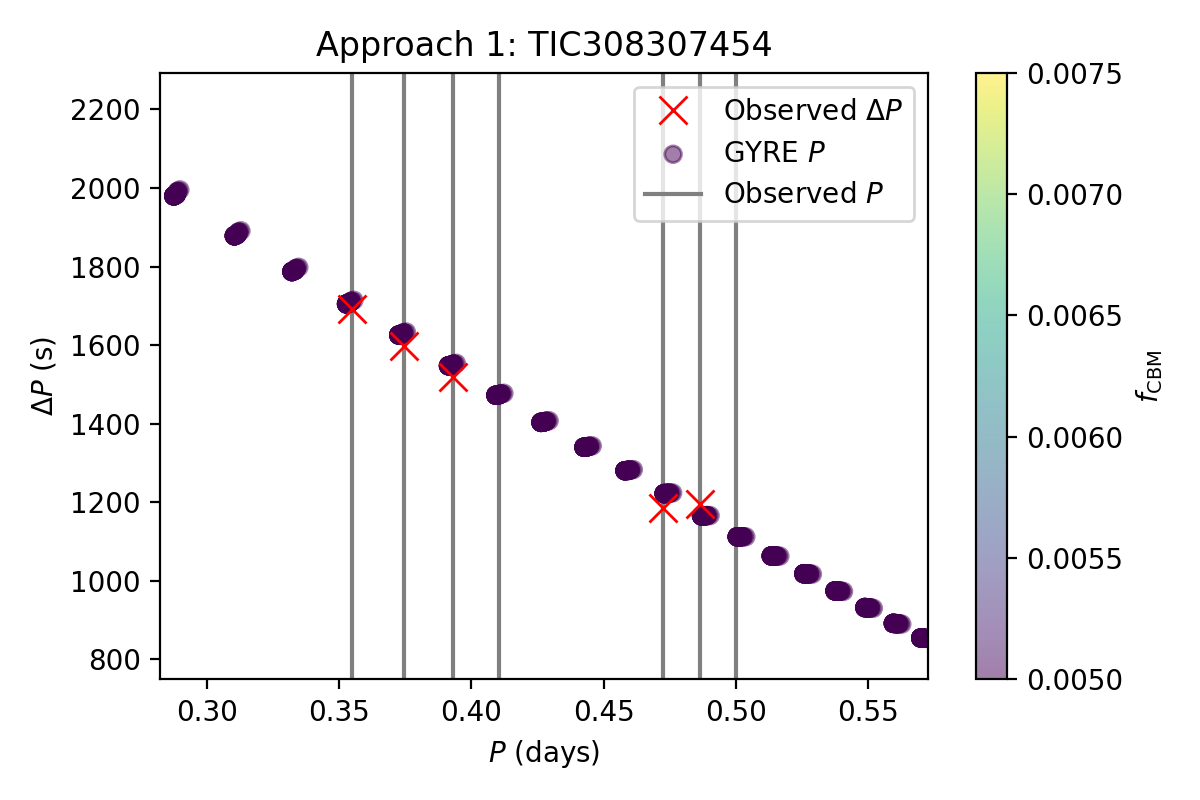}
    \caption{Observed and best-fitting period spacing pattern of TIC\,308307454 from Approach 1. The vertical grey lines indicate the observed g-mode periods, and their period spacings are shown as red crosses. We randomly selected 30 period-spacing patterns from the MCMC result computed with \texttt{GYRE}, shown as circles, with the colour bar indicating the value of $f_\mathrm{CBM}$. }
    \label{fig:DP_approach1_TIC308307454}
\end{figure}

TIC\,308307454 is the star with the lowest mass in our sample. The posterior distributions of its parameters are shown in Fig.~\ref{fig:corner_TIC308307454}, and the best-fitting period spacing pattern is shown in Fig.~\ref{fig:DP_approach1_TIC308307454}. 
Our best-fitting \ca{seismic} model yields a mass of $1.537 \pm 0.008\,\mathrm{M_\odot}$, which lies within $1\sigma$ range of the 
\ca{MIST isochronal} mass. As shown in Fig.~\ref{fig:DP_approach1_TIC308307454}, both the observed and theoretical period spacing patterns are smooth, indicating that 
\ca{dips due to mode trapping caused by a chemical gradient are absent} \citep{Miglio2008MNRAS,Bouabid2013,Pedersen2018}. We obtain a strong constraint on the overshooting parameter $f_\mathrm{CBM}$ as most of the MCMC chains converge to a value of $f_\mathrm{CBM} = 0.005$.

For TIC\,308307454, we computed stellar evolution models with $\log \eta$ values up to 7. However, as shown in Fig.~\ref{fig:corner_TIC308307454}, 
\ca{the star reveals} weak envelope mixing: the posterior distribution of $\log \eta$ is roughly flat below 4 and then drops sharply towards larger values. This result suggests that 
the star with $M \sim 1.5\,\mathrm{M_\odot}$ has low envelope mixing at early evolutionary stages on the main sequence.

The age estimate from Approach~1 is $152 \pm 18\,\mathrm{Myr}$, which is larger than the \texttt{MIST} isochrone age reported by \citet{LiGang_2024_NGC2516}. This star provides a precise age constraint, as the asymptotic spacing in this mass range decreases monotonically with age \ca{\citep{Mombarg2019}.} \citet{LiGang_2024_NGC2516} reported an observed asymptotic spacing of $\Pi_0 = 4340^{+150}_{-160}\,\mathrm{s}$ and a 
near-core rotation rate of $f_\mathrm{rot} = 0.97^{+0.03}_{-0.04}\,\mathrm{d^{-1}}$. 
\ca{These observed values are model independent.}
Our best-fitting model yields a higher value of $\Pi_0$, namely $4504\,\mathrm{s}$, which lies at the upper boundary of the $\pm 1\sigma$ observational range. Although we did not use the \ca{measured} rotation rate as a constraint in the \ca{seismic modelling but rather considered models rotating at 16.7\% of their critical rate for this star, the best-fitting asteroseismic model delivers a near-core rotation rate of $0.973\,\mathrm{d^{-1}}$, consistent with the measured value from the observed period spacing pattern.}

\LGtwo{We also test how different choices of the mixing-length parameter $\alpha_\mathrm{MLT}$ affect our model results, particularly the calculated period-spacing pattern. As shown in Appendix~\ref{app_sec:MLT}, we find that varying $\alpha_\mathrm{MLT}$ has no appreciable impact on the modelling results. }

\subsection{TIC\,281582674}\label{subsec:281582674}
\begin{figure}
    \centering
    \includegraphics[width=0.9\linewidth]{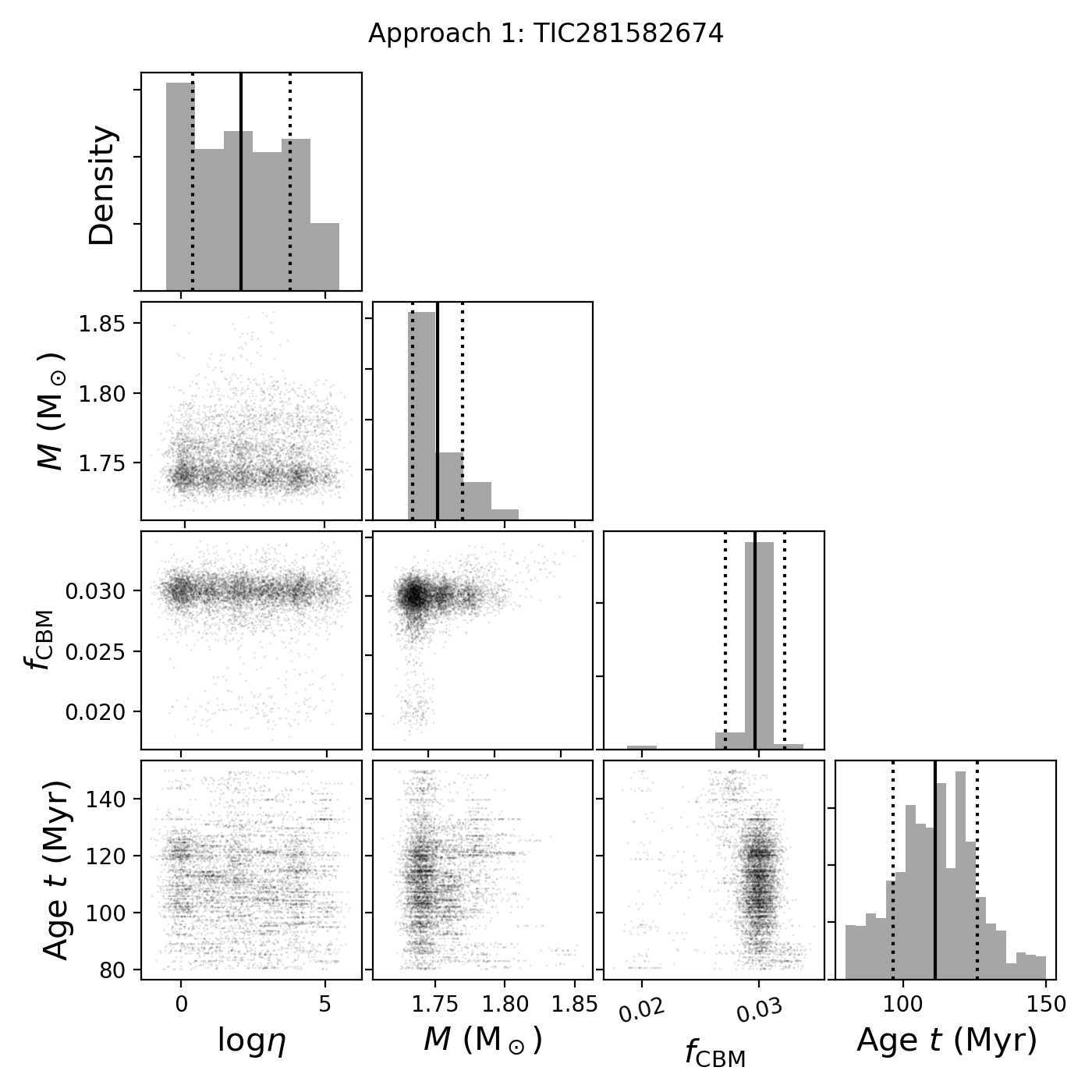}
    \caption{Same as Fig.~\ref{fig:corner_TIC308307454}, but for TIC\,281582674.}
    \label{fig:corner_TIC281582674}
\end{figure}

\begin{figure}
    \centering
    \includegraphics[width=1\linewidth]{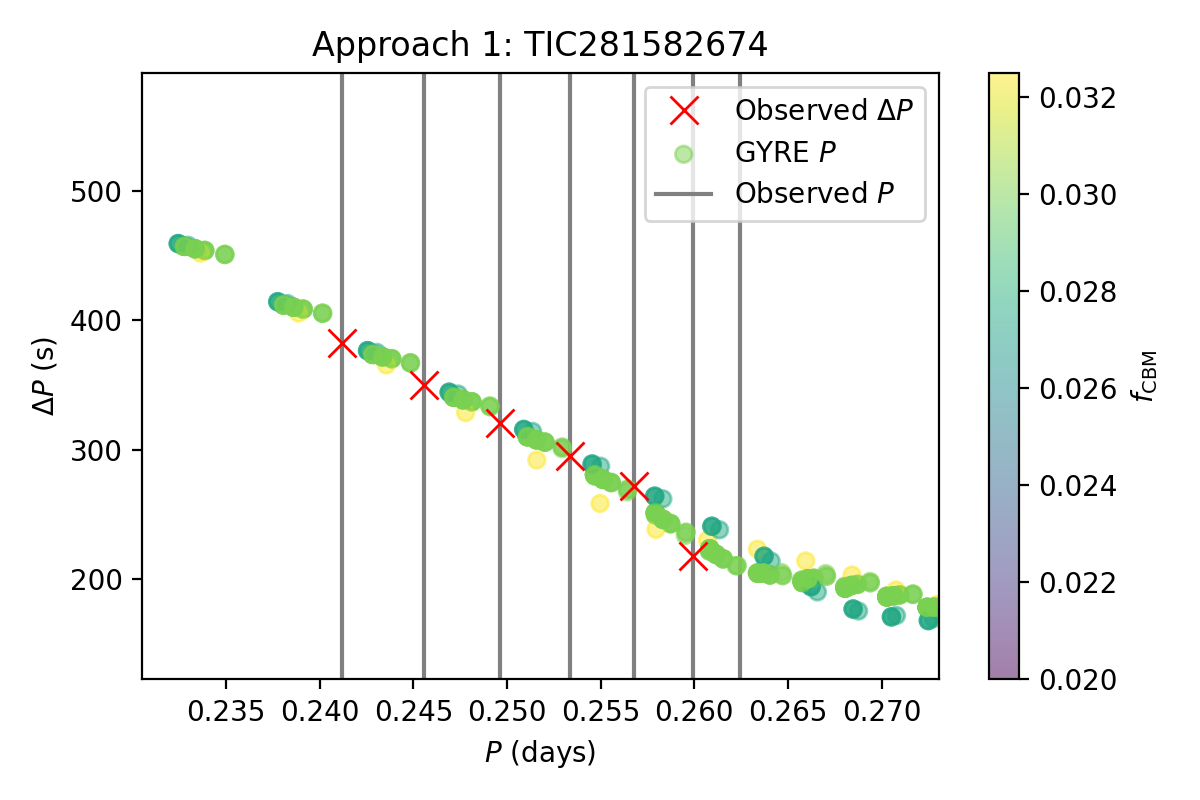}
    \caption{Same as Fig.~\ref{fig:DP_approach1_TIC308307454}, but for TIC\,281582674. }
    \label{fig:DP_approach1_TIC281582674}
\end{figure}

We use TIC\,281582674 as a second representative example, as this star has the highest mass in our sample \ca{of four pulsators}
and is the only one that exhibits a clear dip in its period spacing pattern. The posterior distributions and the best-fitting period spacing pattern are shown in Figs.~\ref{fig:corner_TIC281582674} and \ref{fig:DP_approach1_TIC281582674}, respectively. Our best-fitting \ca{seismic} model yields a mass of $1.752 \pm 0.018\,\mathrm{M_\odot}$, which is $1\sigma$ lower than the isochrone-derived mass. Consequently, the model predicts a lower effective temperature and luminosity, both of which lie near the lower boundary of the $\pm2\sigma$ observational ranges. This lower mass is required to match the observed asymptotic period spacing
\ca{of $4840 \pm 60\,\mathrm{s}$ by \citealt{LiGang_2024_NGC2516}). Our best seismic model has  $\Pi_0=4872$\,s.}

The red crosses in Fig.~\ref{fig:DP_approach1_TIC281582674} represent the observed period spacing pattern. A `half dip' is visible on the long-period side. As discussed previously, we fitted period spacings instead of individual periods, since using period spacing as the observable $\boldsymbol{Y}^\mathrm{obs}$ is more sensitive to such a dip. In the same figure, we find that the location of the dip is highly sensitive to the value of $f_\mathrm{CBM}$. The model with $f_\mathrm{CBM} = 0.03$ best reproduces the observed dip, while models with other values fail to do so. Therefore, this star also places a strong constraint on the overshooting parameter, yielding $f_\mathrm{CBM} = 0.0300 \pm 0.0025$.
We also find that the posterior distribution of $\log \eta$ is similar to that of TIC\,308307454, with a preference for lower values, implying weak envelope mixing in this star as well.

The stellar age is also well constrained in this case. Although the dependence of $\Pi_0$ on age is relatively flat within the age range of NGC\,2516 (see discussion in Section~\ref{subsec:overshoot}), the position of the dip in the period spacing pattern evolves sensitively with age and helps constrain it. We derive an age of $111 \pm 15\,\mathrm{Myr}$ for this star, which is younger than TIC\,308307454 and closer to the \texttt{MIST}-based age.

\LGtwo{To test the sensitivity of the mixing length parameter ($\alpha_\mathrm{MLT}$, fixed to 1.8 in this work) on the calculated oscillation periods, we computed two additional models with $\alpha_\mathrm{MLT}=2.0$ for the best-fitting cases of TIC\,308307454 and TIC\,281582674. We find that varying $\alpha_\mathrm{MLT}$ does not significantly affect the evolutionary tracks, $\Pi_0$, or the g-mode pulsation periods when mode trapping is absent. However, different $\alpha_\mathrm{MLT}$ values do change the g-mode cavity slightly, and consequently alter the time at which the dip in TIC\,281582674 appears. We present these results and further discussion in Appendix~\ref{app_sec:MLT}.}

We also present the posterior distributions and the best-fitting period spacing patterns using Approach~1 for the remaining two stars, TIC\,358466729 and TIC\,410452218, in Figs.~\ref{appen_appro1_fig:corner_TIC358466729}, \ref{appen_appro1_fig:DP_approach1_TIC358466729}, \ref{appen_appro1_fig:corner_TIC410452218}, and \ref{appen_appro1_fig:DP_approach1_TIC410452218} of Appendix~\ref{app_sec:fitting_approach1}. We find that these two stars have 
\ca{more} limited capacity to constrain their parameters, showing 
\ca{somewhat} larger uncertainties compared to the previous two stars. For the fitting results by Approach~2, we present similar figures in Appendix~\ref{app_sec:fitting_approach2}.

\subsection{Overshoot $f_\mathrm{CBM}$}\label{subsec:overshoot}
\begin{figure}
    \centering
    \includegraphics[width=0.85\linewidth]{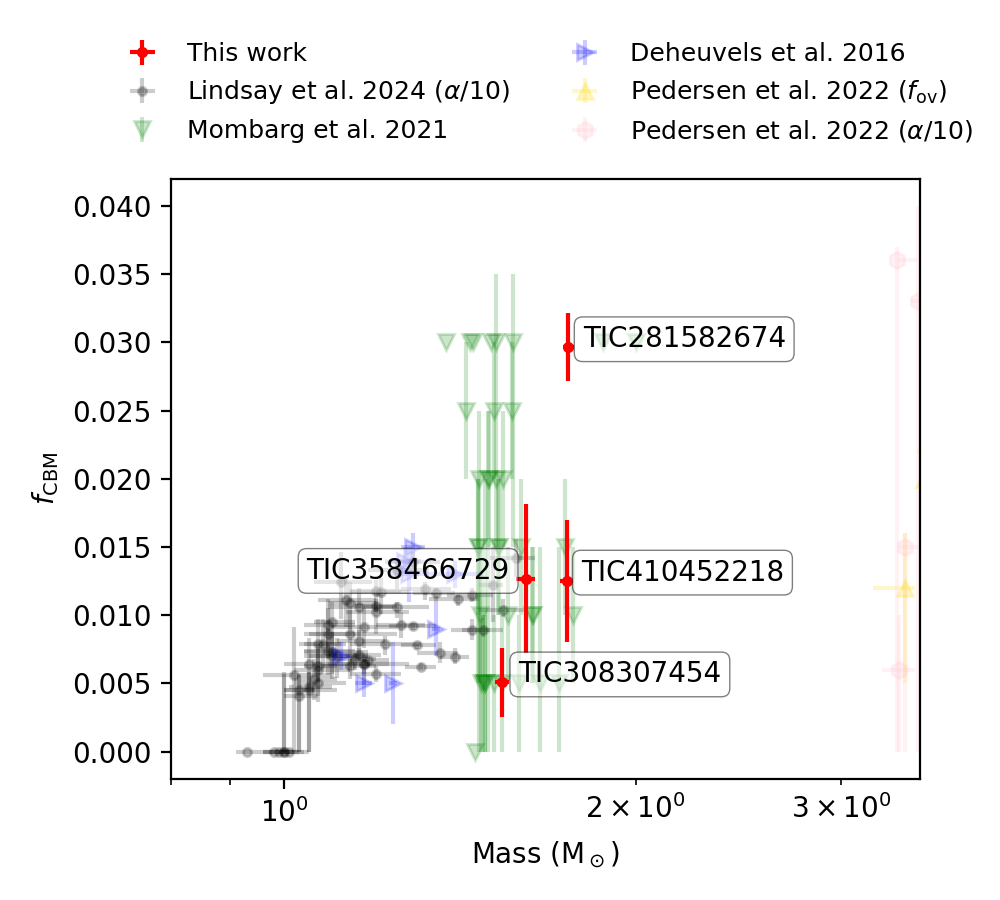}
    \caption{Convective mixing parameter $f_\mathrm{CBM}$ as a function of mass by Approach~1. The red points and their error bars represent the four g-mode pulsators analysed in this work. The translucent markers indicate results from previous studies. }
    \label{fig:overshoot_vs_mass}
\end{figure}

Figure~\ref{fig:overshoot_vs_mass} summarises the overshoot measurements of the four g-mode pulsators in NGC\,2516 fitted in this work using Approach~1.  Apart from the four g-mode pulsators in this work, we also include overshoot measurements from previous studies. For intermediate- to high-mass g-mode pulsators, we show the results of 37 $\gamma$\,Dor stars modelled using a deep-learning method by \citet{Mombarg2021}, and 26 SPB stars from \citet{Pedersen2022}. For solar-like oscillators, we include the fossil overshoot measurements during the main-sequence stage reported by \citet{Deheuvels2016} and \citet{Lindsay2024}. Although these solar-like oscillators have evolved to the post-main-sequence stage, the chemical composition gradients produced by core overshooting remain as fossil signatures and can be detected through mixed modes \citep{Deheuvels2011}. We also note the studies by \citet{Noll2023} and \citet{Noll2021}, which reported overshooting values comparable to those from other works in the mass range below $1.5\,\mathrm{M_\odot}$. In contrast, \citet{Viani2020} reported significantly larger overshooting values, with some stars showing $\alpha_\mathrm{ov}$ as high as 0.4. These discrepancies may stem from differences in the treatment of the temperature gradient within the overshooting region. For this reason, we do not include the results by \citet{Viani2020} in the present comparison.

As listed in Table~\ref{tab:best_fitting_results}, TIC\,308307454 shows the smallest value of $f_\mathrm{CBM}$, while TIC\,281582674 exhibits the largest. The remaining two stars, TIC\,358466729 and TIC\,410452218, display intermediate levels of overshooting. The values of $f_\mathrm{CBM}$ for our g-mode pulsators, which have masses between approximately $1.5$ and $2.0\,\mathrm{M_\odot}$ (corresponding to $\gamma$\,Dor-type pulsators), show a wide spread ranging from 0.005 to 0.030. This spread is consistent with the findings of \citet{Mombarg2021}. A similarly large spread in overshooting parameters has also been reported for SPB stars by \citet{Pedersen2022}, with some values reaching $f_\mathrm{CBM} = 0.040$, or $\alpha_\mathrm{ov} = 0.4$ under the step overshooting scheme. However, we find that the range of $f_\mathrm{CBM}$ values varies with stellar mass. For low-mass stars ($M \lesssim 1.5\,\mathrm{M_\odot}$), $f_\mathrm{CBM}$ is typically smaller than $\sim0.015$, more tightly distributed, and shows an increasing trend with mass. \LG{Furthermore, \cite{Mombarg2024} found that most $\gamma$\,Dor stars have relatively low values of $f_\mathrm{CBM}$, with a linearly decreasing probability density distribution for increasing $f_\mathrm{CBM}$, based on 539 stars.}

We therefore emphasise the importance of $f_\mathrm{CBM}$ in isochrone-based age dating. For young open clusters whose main-sequence turn-off lies within the B- to A-type range, the derived age is primarily constrained by the evolution of these early-type stars. However, such stars show a large spread in the overshoot parameter, which significantly affects stellar evolution, particularly the duration of the main-sequence phase. As a result, adopting a single value of $f_\mathrm{CBM}$ across the entire mass range may not accurately reproduce the true cluster age, although this is a commonly used compromise \citep[such as the \texttt{MIST} isochrone that adopted a fixed value of $f_\mathrm{CBM} = 0.016$ calibrated on the open cluster M67; e.g.][]{Magic2010, Choi2016ApJ_MIST}. An optimised tactic, such as `isochrone-cloud' fitting to the CMD of clusters using multiple combinations of parameters ($f_\mathrm{CBM}$, rotation, radiative envelope mixing, etc.), is needed to assess the age uncertainty arising from stellar models \citep[e.g.][]{Johnston2019A&A,Johnston2021}.

\subsection{Individual seismic ages of the g-mode pulsators in NGC\,2516}

\begin{figure}
    \centering
    \includegraphics[width=\linewidth]{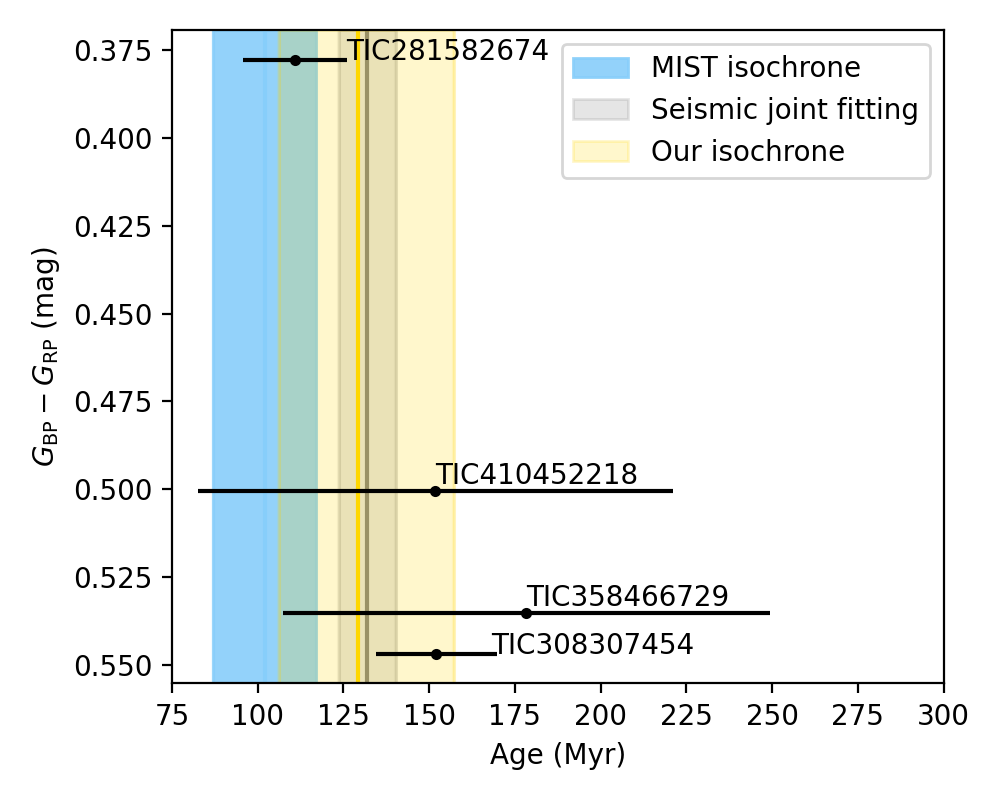}
    \caption{Seismic age estimates for the four g-mode stars in NGC\,2516. Stars are vertically ordered by increasing Gaia colour index $G_{\rm BP} - G_{\rm RP}$. Horizontal error bars indicate the individual seismic ages derived using Approach~1. The vertical grey line and the grey shaded region mark the seismic age and $\pm1\sigma$ range obtained from joint modelling of all the four stars using Approach~2 (“Seismic joint fitting”). The vertical blue shaded region represents the \texttt{MIST} isochrone age derived from CMD fitting. The vertical gold shaded region represents the age derived using our own isochrone \ca{deduced from the seismic model grid}, see Section~\ref{sec:own_isochrone}. }
    \label{fig:age_measure}
\end{figure}

\begin{figure*}
    \centering
    \includegraphics[width=0.9\linewidth]{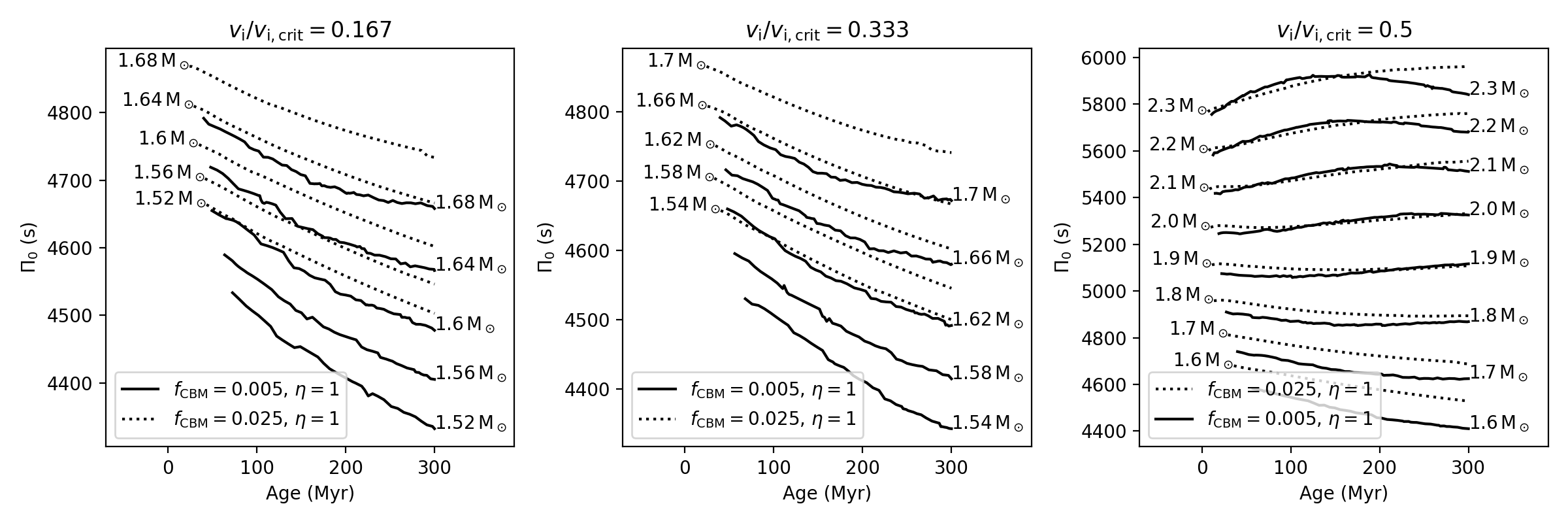}
    \caption{Asymptotic spacing $\Pi_0$ as a function of age, for models with different initial rotation rates and overshooting parameters. The left, middle, and right panels correspond to initial rotation rates of $\Omega/\Omega_{\rm crit,i} = 0.167$, 0.333, and 0.5, respectively. The mixing parameter $\eta$ is fixed at 1 in all models. Each set of curves shows evolutionary tracks for different stellar masses with two overshoot values: $f_{\rm CBM} = 0.005$ (solid lines) and $f_{\rm CBM} = 0.025$ (dashed lines). Masses for the dotted curves ($f_{\rm CBM}=0.025$) are indicated to the left of each line, while those for the solid curves ($f_{\rm CBM}=0.005$) are placed to the right.  }
    \label{fig:Pi0_age}
\end{figure*}

In Fig.~\ref{fig:age_measure}, we summarise the age measurements of the four g-mode stars in NGC\,2516 using both Approach~1 and Approach~2. In Approach~1, where individual ages are allowed for each star, we find a significant spread in the inferred ages. TIC\,281582674 yields the youngest age ($111 \pm 15\,\mathrm{Myr}$), which is consistent with the \texttt{MIST} isochrone age reported by \citet{LiGang_2024_NGC2516}.
In contrast, TIC\,308307454 has an age of $152 \pm 18\,\mathrm{Myr}$, which deviates from the \texttt{MIST} isochrone age at the $2\sigma$ level.
We note that TIC\,410452218 and TIC\,358466729 exhibit large uncertainties in their age estimates, implying a limited capacity for precise age determination. 

It is well known that the asymptotic period spacing, $\Pi_0$, serves as an age indicator in g-mode pulsators \citep[e.g.][]{Miglio2008MNRAS, Saio2018, Ouazzani2019A&A, Mombarg2019}, as $\Pi_0$ decreases monotonically with age during the main-sequence phase. To explore the relationship between $\Pi_0$ and age in our modelling, we plot $\Pi_0$ as a function of age in Fig.~\ref{fig:Pi0_age}, adopting a fixed value of $\eta=1$ and considering two values of $f_\mathrm{CBM}$. 
We find that the evolutionary trend of $\Pi_0$ with age varies significantly with stellar mass. For stars around $1.5\,\mathrm{M_\odot}$, $\Pi_0$ exhibits a sharp decline over time, with a decrease of up to $200\,\mathrm{s}$ within $300\,\mathrm{Myr}$—exceeding the typical observational uncertainty of $\Pi_0$ (approximately $100\,\mathrm{s}$). This explains why TIC\,308307454 ($M \sim 1.54\,\mathrm{M_\odot}$) can provide a well-constrained age. However, the sensitivity of $\Pi_0$ to age diminishes with increasing mass: at $M \sim 1.7\,\mathrm{M_\odot}$, the total variation in $\Pi_0$ is roughly $100\,\mathrm{s}$ within 300\,Myr, while at $M \sim 1.8\,\mathrm{M_\odot}$, $\Pi_0$ remains nearly constant. This mass range corresponds to TIC\,358466729 and TIC\,410452218, which explains why these two stars yield \ca{somewhat poorer} constrained ages. TIC\,281582674, on the other hand, benefits from the presence of a dip in its period spacing pattern, which aids in constraining its age. We therefore emphasise that precise age constraints from g-mode pulsators require either low-mass stars (such as $\gamma$\,Dor stars) or the detection of mode trapping signatures in the period spacing patterns.

We also find that $\Pi_0$ starts to increase at the beginning of the main sequence if the mass is approximately greater than $1.8\,\mathrm{M_\odot}$. The behaviour may change depending on the adopted value of $f_\mathrm{CBM}$. In the right panel of Fig.~\ref{fig:Pi0_age} for $v_\mathrm{i} /v_\mathrm{crit, i} = 0.5$, we find that $\Pi_0$ first increases and then decreases for $f_\mathrm{CBM} = 0.005$, while it continues to increase for $f_\mathrm{CBM} = 0.025$. These features may challenge age determinations using high-mass g-mode pulsators in young open clusters, as $\Pi_0$ no longer shows a monotonic relation with age. An increasing $\Pi_0$ is indicative of a growing convective core \citep[e.g.][]{Mombarg2019, Aerts2025}. As simulated by \citet{Temaj2024}, a larger $f_\mathrm{CBM}$ leads the core to grow in mass, causing the star to behave as if it had a higher initial mass and ultimately altering its final fate.

In addition to modifying whether $\Pi_0$ increases or decreases with age, we find that $f_\mathrm{CBM}$ also affects the absolute value of $\Pi_0$. When the mass is roughly below $1.8\,\mathrm{M_\odot}$, $\Pi_0$ always decreases monotonically with age, but smaller $f_\mathrm{CBM}$ values tend to produce lower $\Pi_0$ values. The reduction can be as large as $100\,\mathrm{s}$ and gradually disappears when $M \gtrsim 1.8\,\mathrm{M_\odot}$.

\subsection{Joint age measurement of NGC\,2516}\label{subsec:joint_age}

As introduced in Section~\ref{subsec:individual_or_joint_fitting}, instead of allowing the stars to have different ages (Approach~1), we assumed that the stars share a common age, as they were born simultaneously in the same cluster.
\ca{We thus} fitted them jointly using the likelihood defined in Eq.~\ref{eq:joint_likelihood} (Approach~2). This joint fitting approach yields an asteroseismology-derived age for NGC\,2516 of \jointage.

The vertical grey band in Fig.~\ref{fig:age_measure} represents the joint age result. We find that the seismic joint age (\jointage) differs from the \texttt{MIST} isochrone age (\MISTage) by approximately $1.5\sigma$. In Section~\ref{sec:own_isochrone}, we demonstrate that this age discrepancy arises from differences in the adopted input physics of the \texttt{MIST} isochrone. When using identical input physics \ca{to define isochrones and perform forward asteroseismic modelling}, the isochrone-fitting and seismic-fitting methods yield consistent age measurements.

The best-fitting period spacing patterns and the posterior distributions from Approach~2 are presented in Appendix~\ref{app_sec:fitting_approach2}. Compared to Approach~1, the period spacing patterns obtained from Approach~2 show a slightly poorer fit, as the condition of a joint age may not optimally reproduce the individual stars' period spacing patterns. Nevertheless, the resulting best-fitting parameters ($M$, $f_\mathrm{CBM}$, $\log \eta$) remain highly consistent with those derived from Approach~1, as listed in Table~\ref{tab:best_fitting_results}


\subsection{\ca{Seismic and \texttt{MIST}-based} mass discrepancy}\label{subsec:mass_discrepancy}
\begin{figure}
    \centering
    \includegraphics[width=0.9\linewidth]{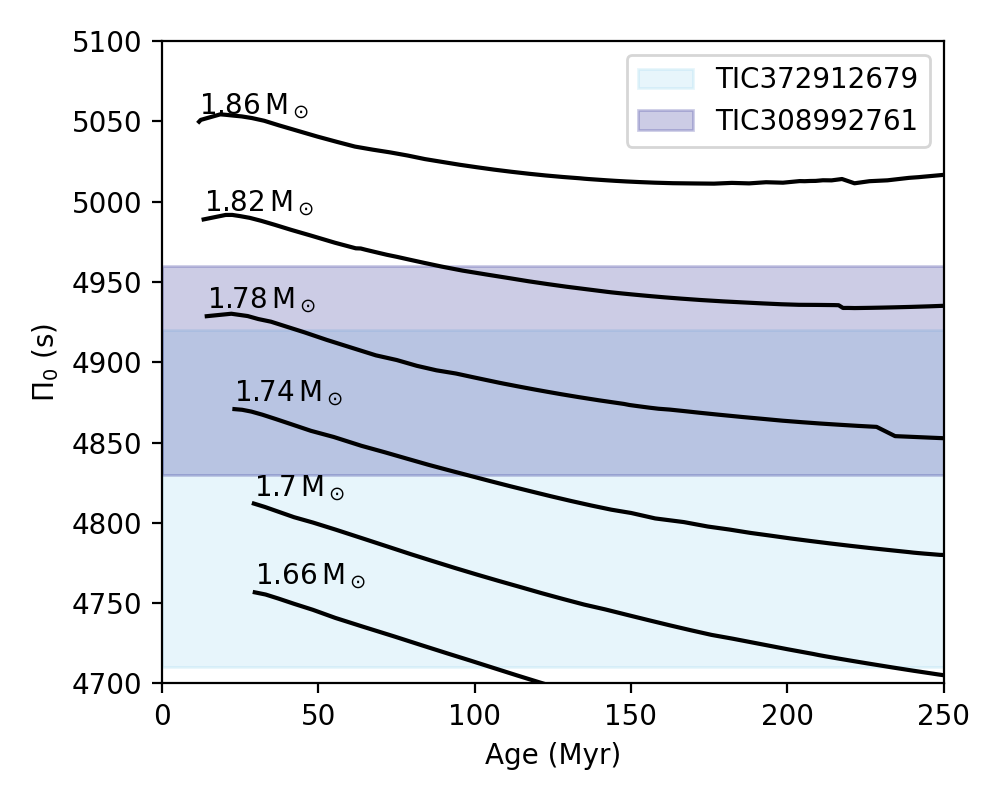}
    \caption{Asymptotic period spacing $\Pi_0$ as a function of stellar age for models with fixed initial rotation $v_\mathrm{i}/v_\mathrm{i,init}=0.5$, $f_\mathrm{CBM}=0.025$, $\log \eta = 3$, but different initial masses. The horizontal shaded regions indicate the observed $\Pi_0$ values (and uncertainties) for TIC\,372912679 and TIC\,308992761. }
    \label{fig:Pi0_mass_discrepancy}
\end{figure}

As mentioned earlier, the two high-mass stars, TIC\,372912679 and TIC\,308992761, could not be modelled. We find that this is because the observed values of $\Pi_0$ cannot be reproduced within the \ca{\texttt{MIST}-based} mass ranges given in Table~\ref{tab:observational_constraints}. Figure~\ref{fig:Pi0_mass_discrepancy} illustrates this situation. As reported by \citet{LiGang_2024_NGC2516}, TIC\,372912679 has $\Pi_0 = 4820^{+100}_{-110}\,\mathrm{s}$, while TIC\,308992761 has $\Pi_0 = 4900^{+60}_{-70}\,\mathrm{s}$ (shown as the horizontal shaded regions in Fig.~\ref{fig:Pi0_mass_discrepancy}, respectively). However, comparison with our 
\ca{seismic} models shows that these $\Pi_0$ values correspond to masses around $1.74\,\mathrm{M_\odot}$, deviating by 2 to 3$\sigma$ from the 
\ca{\texttt{MIST}-derived} masses listed in Table~\ref{tab:observational_constraints} (which are around $2\,\mathrm{M_\mathrm{\odot}}$). We also note that TIC\,281582674 reaches its lower bound of its allowed mass range during the fitting process. Therefore, we find a discrepancy between \ca{our asteroseismic masses and the \texttt{MIST}-based isochrone masses}, and this discrepancy becomes more pronounced with increasing stellar mass.

Mass discrepancies among different stellar mass estimation methods are not rare and have been noted by numerous previous studies. These studies commonly compare evolutionary masses \ca{based on models with one particular choice for the input physics}
with dynamical masses in binary systems.
\ca{Such dynamical masses are derived from the binary orbital motion and are therefore model-independent, making binaries ideal as benchmarks.} For example, a significant mass discrepancy in 
\ca{the evolved massive binary}
V380\,Cyg was reported by comparing its dynamical mass and evolutionary mass inferred from theoretical evolutionary tracks \citep{Guinan2000, Tkachenko2014}. Analyses of binary star samples by \citet{Schneider2014}, \citet{Claret2019}, and \citet{Tkachenko2020} revealed \ca{systematically occurring} mass discrepancies between dynamical and evolutionary masses. These studies argued that both enhanced core overshooting and proper treatment of microturbulent velocities and turbulent pressure in stellar atmosphere models can \ca{partially} resolve the discrepancies.

Pulsating binaries allow us to compare the seismic and dynamical masses. \citet{Schmid2015_10080943_obs} reported the dynamical masses for the two g-mode pulsators in the \ca{non-eclipsing binary KIC\,10080943, which shows a periodic brightening effect along its 15.3\,d eccentric orbit. The masses of the 
primary and secondary were deduced from photometric light curve modelling and spectral disentangling, leading to component} masses of $2.0\pm0.1\,\mathrm{M_\odot}$ and $1.9\pm0.1\,\mathrm{M_\odot}$, respectively, similar to our targets TIC\,372912679 and TIC\,308992761. In a subsequent asteroseismic modelling study, \citet{Schmid2016_10080943_modelling} found that the seismic masses were about $2\sigma$ lower than the \ca{binary masses, hence} their results \ca{for the two $\gamma\,$Dor stars in this close binary} are similar to ours \ca{for the two more massive $\gamma\,$Dor stars in the cluster.}
\citet{Schmid2016_10080943_modelling} attempted to modify the initial helium mass fraction but found that this adjustment did not significantly increase the seismic masses. 

Isochrone-derived masses \ca{close to the regime of the $\gamma\,$Dor stars} have also \ca{been scrutinised.} The eclipsing binary WOCS\,11028 \ca{in M67
was found to show a significant discrepancy between its dynamical and isochrone-derived masses \citep{Sandquist2021}.} Subsequent studies confirmed this discrepancy using updated isochrone models \citep{Nguyen2022, Reyes2024_improved_isochrone, Reyes2025MNRAS}.

Mass \ca{discrepancies are} also found in other mass ranges or in different evolutionary stages, such as O-type stars \citep{Weidner2010, Markova2018}, Cepheids \citep{Neilson2011}, and white dwarfs \citep{Calcaferro2024}. Various attempts have been made to resolve the mass discrepancies by adjusting stellar model parameters, for example, by adopting a larger overshooting parameter to produce a more extended convective core \citep{Tkachenko2020}. However, as shown in Fig.~\ref{fig:Pi0_age}, varying $f_\mathrm{CBM}$ has only a minor effect on $\Pi_0$ in this mass range, \ca{pointing out that the input physics itself needs to be changed, rather than simply adapting the free parameters in the particularly frozen convective core overshooting description. Motivated to resolve the mass discrepancy for the two massive $\gamma\,$Dor pulsators in NGC\,2516, we embarked upon}
self-constructed \ca{asteroseismically-calibrated isochrones. In the next section, we} demonstrate that adopting identical input physics to that used in the seismic models can partially alleviate the mass discrepancy.

\section{Self-constructed isochrone using \ca{input physics from asteroseismology}}\label{sec:own_isochrone}

\begin{figure*}
    \centering
    \includegraphics[width=0.9\linewidth]{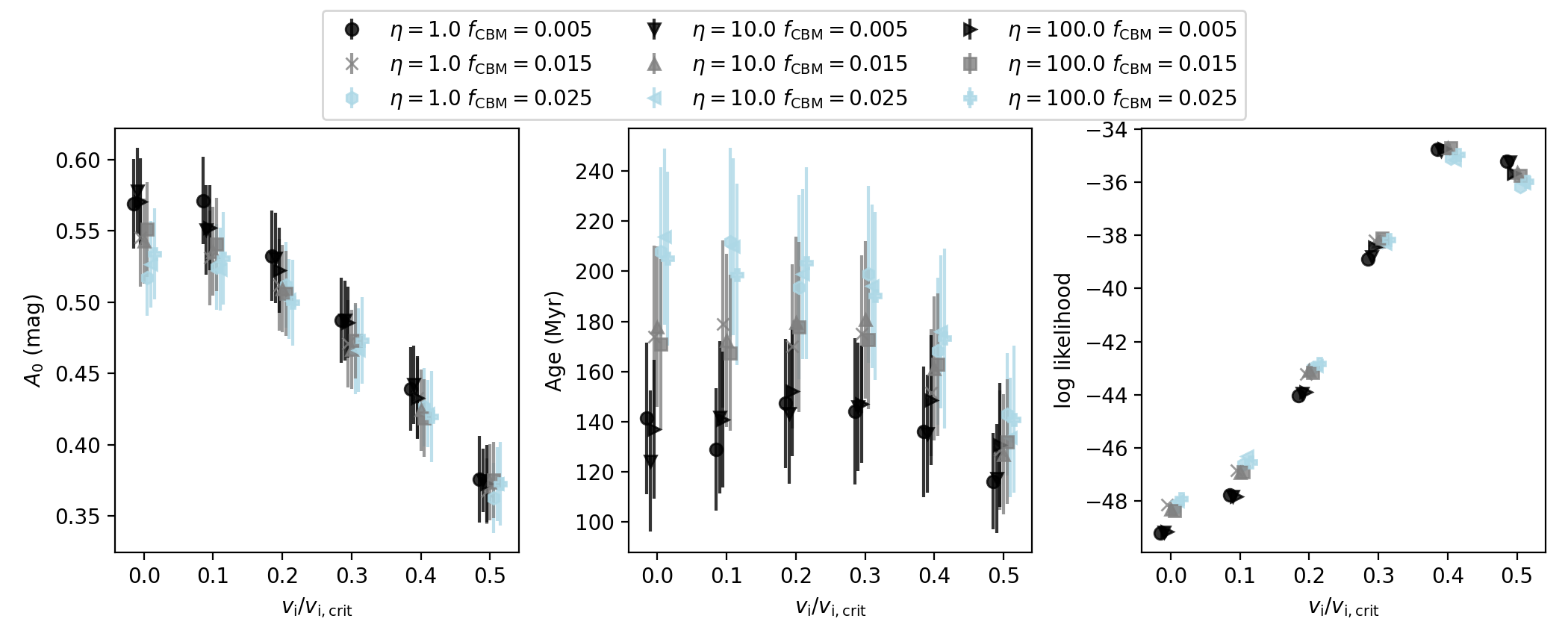}
    \caption{Results of our self-constructed isochrone fitting for various combinations of convective boundary mixing and envelope mixing efficiency, as a function of initial rotation fraction $v_\mathrm{i}/v_\mathrm{i,crit}$. The legend at the top indicates the different combinations of $\eta$ (envelope mixing parameter) and $f_\mathrm{CBM}$ (convective boundary mixing). Symbols of the same colour share the same value of $f_\mathrm{CBM}$, while different colours correspond to different values of $\eta$. The data points are artificially shifted to avoid overlapping. Left: the Gaia extinction parameter $A_0$ as a function of initial rotation fraction. Middle: the best-fitting cluster age as a function of rotation. Right: the log-likelihood of the best-fitting isochrone. Error bars indicate 1$\sigma$ confidence intervals.}
    \label{fig:self_isochrone_results}
\end{figure*}


In Section~\ref{subsec:joint_age}, we reported that the joint seismic age of NGC\,2516 deviates from the \texttt{MIST} isochrone age by $2\sigma$. In Section~\ref{subsec:mass_discrepancy}, we also discovered mass discrepancies for two high-mass g-mode pulsators when compared to the \texttt{MIST} masses. Given that the \texttt{MIST} isochrones are constructed using different input physics, which may introduce inconsistencies with our seismic modelling, we therefore construct our own isochrones. 
\ca{These} are both physically consistent with and calibrated by asteroseismology. 
\ca{Subsequently we assess whether the discrepancies are resolved by the asteroseismic isochrone of NGC\,2516.}

\subsection{Construction of our own isochrones}

To simplify the construction procedure, \ca{we customise our isochrone development to} young open clusters. We consider only the upper main sequence of the CMD as input observables, corresponding to main-sequence stars with masses larger than $\sim\!1.5\,\mathrm{M_\odot}$
\ca{(see black dots in Fig.~\ref{fig:best_fitting_self_construted_isochrone} discussed below). }
This mass boundary can be visually identified from the CMD, where the slope of the main sequence changes noticeably. We used the member stars of NGC\,2516 reported by \cite{Meingast2021} and discarded both low-mass main-sequence stars and post-main-sequence stars in NGC\,2516. \ca{This simplifies the isochrone construction} by avoiding the need to consider different physical treatments required for cool stars and evolved stars. We checked the \texttt{Gaia} \texttt{ruwe} values for the member stars and excluded those with $\mathrm{ruwe} > 1.4$, which may indicate binarity. However, only a few stars were affected. In addition, two potential blue stragglers were excluded from the fitting process. 

As stars on the main sequence evolve slowly, we adopted a mass step of $0.05\,\mathrm{M_\odot}$, which is sufficiently fine to reproduce the structure near the main-sequence turn-off. We computed evolutionary tracks from $1.5\,\mathrm{M_\odot}$ to $8\,\mathrm{M_\odot}$, covering stellar ages from a few million years up to several hundred million years. \ca{We adopt the same input physics as for the asteroseismology and consider a} range of input \ca{parameters}, including initial rotation rate fractions $v_\mathrm{i}/v_\mathrm{i, crit}$ from 0 to 0.5 in steps of 0.1, convective boundary mixing values of $f_\mathrm{CBM} = 0.005$, 0.015, and 0.025, and envelope mixing parameters $\eta = 1$, 10, and 100.

We briefly summarise the input physics used for our seismic modelling in Section~\ref{susec:input_physics}, \ca{to highlight} the differences with the \ca{input physics of the} \texttt{MIST} isochrones \citep{Choi2016ApJ_MIST}. 
First, we treat angular momentum (AM) transport using a scheme with an asteroseismology-calibrated constant viscosity, and model chemical element diffusion with a differential-rotation-induced mechanism encompassing a diffusion coefficient $\eta$ (Eq.~\ref{eq:rotation_diffusion}), whereas the \texttt{MIST} isochrones adopt \texttt{MESA}'s built-in prescriptions for both AM transport and chemical diffusion \citep[see][for details]{Choi2016ApJ_MIST, Mombarg2023calibrating_AM}. 
Second, \ca{while the formulation of core overshooting is the same,}
we \ca{allow} three values for the exponential overshooting parameter \ca{reflecting the asteroseismic modelling results}, while the \texttt{MIST} models use a fixed value of $f_\mathrm{ov, core} = 0.0160$. 
Third, we adopted a constant initial rotation fraction for all main-sequence stars, set to $v_\mathrm{i}/v_\mathrm{i, crit} = 0.5$. This value was measured from the g-mode period spacing patterns of 11 g-mode pulsators by \citet{LiGang_2024_NGC2516}. In contrast, the \texttt{MIST} isochrones adopt a mass-dependent rotation prescription, where the rotation rate increases linearly from zero at $1.2\,\mathrm{M_\odot}$ to a fixed value at $1.8\,\mathrm{M_\odot}$. 
Fourth, we employ an asteroseismology-calibrated primordial helium abundance and element enrichment rate, whereas the \texttt{MIST} models use cosmology-calibrated ones \citep{PlanckCollaboration2016}.

After obtaining the theoretical evolutionary tracks, which provide the effective temperature and luminosity at each evolutionary time step, we constructed an isochrone by selecting the points on each track corresponding to a given age. We applied bolometric and extinction corrections to transform the theoretical $T_\mathrm{eff}$ and $L$ into observed magnitudes and colour indices. To achieve this, we adopted the same bolometric corrections as the \texttt{MIST} isochrones, which are based on a grid of stellar atmospheres and synthetic spectra computed using the ATLAS12 and SYNTHE codes \citep{Kurucz2005, Choi2016ApJ_MIST}. We adopted the \texttt{Gaia} extinction law given for the \texttt{Gaia} photometric bands\footnote{\url{https://www.cosmos.esa.int/web/gaia/edr3-extinction-law}} \citep{Danielski2018, Fitzpatrick_2019, Riello2021}.

\subsection{Fitting process}
As we only used the upper main sequence as the input observables, the shape of the CMD was relatively simple, exhibiting a monotonic relation between the absolute G-band magnitude and the \texttt{Gaia} colour index $G_\mathrm{BP}-G_\mathrm{RP}$. Therefore, we considered the absolute G-band magnitude as the independent variable and minimised the residuals of the \texttt{Gaia} colour index to determine the best-fitting model. The uncertainty of the \texttt{Gaia} colour index was estimated using the same method as in Section~\ref{sec:sample_obs_constraints}. This uncertainty was significantly larger than the photometric error bars provided by the \texttt{Gaia} archive, as it contains additional sources of uncertainty beyond pure photometric noise, as discussed in Section~\ref{sec:sample_obs_constraints}. An MCMC algorithm was used to maximise the likelihood function based on the residuals of the \texttt{Gaia} colour index. Flat priors for $A_0$ and age were applied. During the CMD fitting, we treated the extinction at $\lambda = 550\,\mathrm{nm}$ ($A_0$) and the cluster age as two free parameters. The observed CMD was fitted using all combinations of $f_\mathrm{CBM}$, $\eta$, and $v_\mathrm{i}/v_\mathrm{i, crit}$. 

\subsection{Results of the self-constructed isochrones}
The best-fitting results are shown in Fig.~\ref{fig:self_isochrone_results}. The left panel displays the derived values of $A_0$ for various combinations of input parameters. We find that the rotation rate has the dominant effect on $A_0$: faster rotation leads to significantly lower values of $A_0$, whereas the other two parameters ($f_\mathrm{CBM}$ and $\eta$) have negligible influence. This is because rotation substantially changes the stellar structure and surface temperature distribution, thereby affecting the extinction.

In the middle panel of Fig.~\ref{fig:self_isochrone_results}, we find that both $f_\mathrm{CBM}$ and rotation influence the age determination. Larger overshooting results in older inferred ages, while faster rotation leads to younger ages. Additionally, the age estimates converge as $v_\mathrm{i}/v_\mathrm{i, crit}$ increases. The right panel of Fig.~\ref{fig:self_isochrone_results} shows the likelihood values of the best-fitting isochrones. We find that isochrones with higher rotation rates generally provide better fits. We also noted that the isochrone fitting quality is insensitive to the values of 
$\eta$. Upon inspecting the best-fitting isochrones overlaid on the observed CMD, we realise that isochrones with low rotation rates fail to reproduce the observed main sequence, resulting in lower likelihood values.

Based on both the likelihood function and the asteroseismic results, we selected the isochrones with $v_\mathrm{i}/v_\mathrm{i, crit} = 0.5$ as the best-fitting models. However, the inferred age also exhibits a slight dependence on the choice of $f_\mathrm{CBM}$, ranging from approximately $120\,\mathrm{Myr}$ to $140\,\mathrm{Myr}$ as $f_\mathrm{CBM}$ increases from 0.005 to 0.025. Considering the asteroseismic measurements of the overshooting parameter in B-type stars \citep[as shown in Fig.~\ref{fig:overshoot_vs_mass},][]{Pedersen2022}, which suggest a median value of $f_\mathrm{CBM} \approx 0.025$, we combined the posterior distributions with all $\eta=1,~10,~100$ and $f_\mathrm{CBM}=0.005,~0.015,~0.025$ as final posterior distributions. This strategy is a solution to the so-called `isochrone-cloud' fitting, as it takes the 
\ca{occurring diversity in mixing levels} into account \citep[e.g.][]{Johnston2019A&A,Johnston2021}. 
Finally, we obtain $A_0 = 0.37 \pm 0.03\,\mathrm{mag}$ and an age of \selfisochroneage. 
As a reminder, the \texttt{MIST} fit yielded $A_{0,\mathrm{MIST}} = 0.53 \pm 0.04\,\mathrm{mag}$ and an age of \MISTage~\citep{LiGang_2024_NGC2516}. 
We show the best-fitting self-constructed isochrone in Fig.~\ref{fig:best_fitting_self_construted_isochrone}. 

We find that, compared to the previous \texttt{MIST} $A_0$ measurement, the newly derived extinction, $A_0 = 0.37 \pm 0.03\,\mathrm{mag}$, agrees better with other independent extinction estimates. Using the \texttt{Gaia} extinction law, we convert $A_0$ to reddening at $8000\,\mathrm{K}$, yielding $E\left(\mathrm{BP{-}RP}\right) = 0.18 \pm 0.04\,\mathrm{mag}$. Applying the conversion coefficient from \citet{Casagrande2018}, which states that $E\left(\mathrm{BP{-}RP}\right) = \left(3.374 - 2.035\right) E\left(\mathrm{B{-}V}\right)$, we derive a corresponding $E\left(\mathrm{B{-}V}\right) = 0.13 \pm 0.03\,\mathrm{mag}$. This revised reddening is more consistent with the result of \citet{Sung2002AJ}, who reported $E\left(\mathrm{B{-}V}\right) = 0.112 \pm 0.024\,\mathrm{mag}$, and also agrees better with the \texttt{Gaia} total Galactic extinction map\footnote{\url{https://gea.esac.esa.int/archive/documentation/GDR3/Data_analysis/chap_cu8par/sec_cu8par_apsis/ssec_cu8par_apsis_tge.html}} \citep[which gave $A_0 = 0.4358$;][]{Delchambre2023}.

The newly derived age from our self-constructed isochrone is in good agreement with the seismic age. As shown in Fig.~\ref{fig:age_measure}, the age inferred from the self-constructed isochrone has nearly the same median value as the seismic age, although it \ca{has} a larger uncertainty. This is expected, as red giants were not included in the isochrone fitting and the main-sequence turn-off exhibits significant scatter. We thus demonstrate that the asteroseismology-calibrated isochrone yields an age for NGC\,2516 that is consistent with the result from seismic modelling \ca{of four g-mode pulsators in the cluster.}

\begin{figure}
    \centering
    \includegraphics[width=0.85\linewidth]{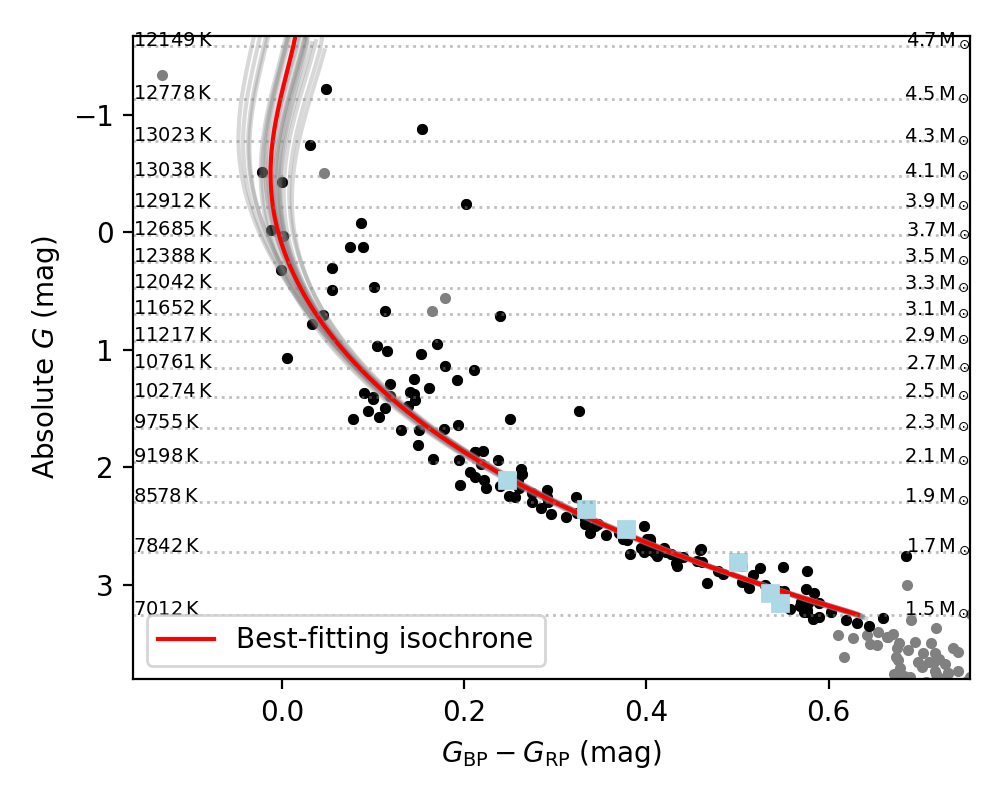}
    \caption{Best-fitting result of the self-constructed isochrone with $\eta = 10$, $f_\mathrm{CBM} = 0.025$, and $v_\mathrm{i}/v_\mathrm{i, crit} = 0.5$. The red curve represents the best-fitting isochrone, while the grey curves show 20 randomly selected samples from the MCMC chains, showing the uncertainty region. The small grey dots indicate member stars from \citet{Meingast2021}, and the black dots are the upper main-sequence stars used as input observables.  The modelled g-mode pulsators are marked with light blue squares.}
    \label{fig:best_fitting_self_construted_isochrone}
\end{figure}

\begin{figure}
    \centering
    \includegraphics[width=0.85\linewidth]{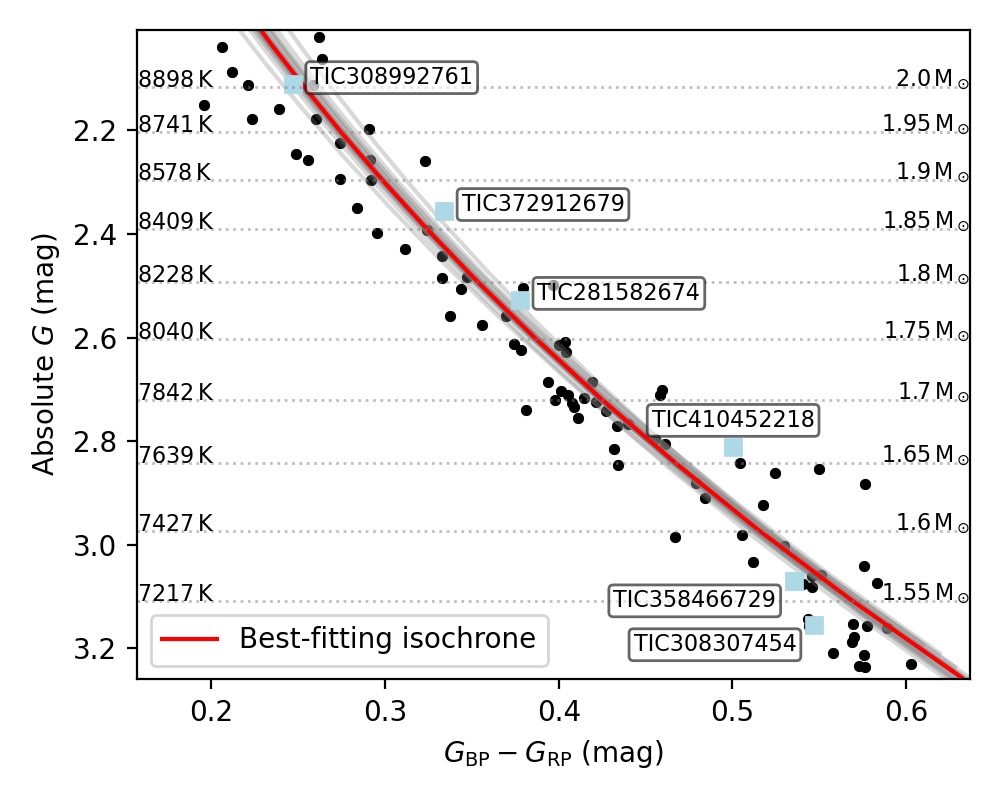}
    \caption{Same as Fig.~\ref{fig:best_fitting_self_construted_isochrone}, but only showing stars with masses up to $2.0\,\mathrm{M_\odot}$. The modelled g-mode pulsators are marked with light blue squares.}
    \label{fig:best_fitting_self_construted_isochrone_zoomin}
\end{figure}

The mass discrepancy is partially, but not entirely, resolved when using our self-constructed isochrone with asteroseismology-measured rotation rates. The revised mass estimates from the best-fitting self-constructed isochrone for the six g-mode pulsators are shown in Fig.~\ref{fig:best_fitting_self_construted_isochrone_zoomin}. Compared to the \texttt{MIST}-based masses listed in Table~\ref{tab:observational_constraints}, the masses derived from the self-constructed isochrone are systematically lower. For the two lower-mass stars, TIC\,308307454 and TIC\,358466729, the mass differences are small (approximately $-0.04\,\mathrm{M_\odot}$) and are within the error bars. This explains why no significant mass discrepancy \ca{was observed for those pulsators in the first place.} However, the mass differences increase with stellar mass: TIC\,372912679 shows a shift of $-0.08\,\mathrm{M_\odot}$, and TIC\,308992761 shows a shift of $-0.11\,\mathrm{M_\odot}$. We attribute these mass shifts to the adopted rotation rate, as we find that using a self-constructed isochrone with a lower initial rotation rate reduces the mass shifts. This result highlights the importance of carefully selecting initial rotation rates, as they directly influence mass estimates of cluster members. 

Even after accounting for these mass shifts, the self-constructed isochrone still does not fully reconcile the mass values with those obtained from seismic estimation based on \ca{the observed internal structure quantity}
$\Pi_0$ (approximately $1.75\,\mathrm{M_\odot}$; see Fig.~\ref{fig:Pi0_age}), leaving a \ca{small} residual mass discrepancy of \ca{about} $0.1\,\mathrm{M_\odot}$. Such a discrepancy \ca{is similar to the } previous findings in the g-mode binary KIC\,10080943 \citep{Schmid2015_10080943_obs, Schmid2016_10080943_modelling} and highlights a mismatch between the internal and surface \ca{input physics adopted in the best} current one-dimensional stellar models \ca{and the reality of fast-rotating three-dimensional stars}.

\section{Conclusions}\label{sec:conclusions}

In this paper, we presented the first asteroseismic modelling of \ca{a sample of} g-mode main-sequence pulsators in \ca{a} young open cluster. Our approach was to constrain the stellar surface parameters using the best-fitting isochrone in the colour–magnitude diagram (CMD), without relying on any spectroscopic data. This was then followed by seismic modelling based on \ca{observed g-mode period spacing patterns
of $\gamma\,$Dor pulsators in the cluster. Their asteroseismic observables
probe the internal rotation and physical conditions in the transition zone between the convective core and the radiative envelope. These two steps together} provide a novel method to measure cluster ages 
\ca{with high precision} and offer new insights into the internal structure of 
\ca{intermediate-mass stars}, particularly the connection between their cores and surfaces.

Building on the results of \citet{LiGang_2024_NGC2516}, we constrained the effective temperatures, luminosities, and masses of six g-mode pulsators in NGC\,2516, as listed in Table~\ref{tab:observational_constraints}. These tight constraints enabled the construction of a fine \ca{grid of stellar models}. We employed rotating stellar models computed with \texttt{MESA}, incorporating an asteroseismology-calibrated viscosity for angular momentum transport and rotation-induced mixing in the radiative envelope. 
\ca{Theoretical} g-mode pulsation periods were subsequently calculated using \texttt{GYRE}. The best-fitting models were determined by matching the observed and calculated  periods
\ca{of identified g modes}.
We employed two approaches \ca{for the asteroseismic modelling}. In Approach~1, we fitted the stars individually, allowing each to have its own age. In Approach~2, we performed a joint fit, assuming that all stars share a common age. 

Using Approach~1, we determined the stellar masses and the convective core overshooting parameter. The envelope mixing parameter $\eta$ could not be constrained, as the observed smooth period spacings are insensitive to the envelope mixing, pointing to the absence of strong chemical composition gradients. The inferred ages range from approximately $111$ to $152$\,Myr, with some estimates exhibiting large uncertainties due to the slow evolution of $\Pi_0$ near the zero-age main sequence.

Using \ca{the full power of joint modelling of g-mode pulsators in a cluster (Approach~2)}, we determined 
\ca{the high-precision} seismic age for NGC\,2516 \ca{to be} \jointage. This value deviates from the \texttt{MIST}-based age (\MISTage) \ca{at the level of} $2\sigma$. This age discrepancy \ca{points to} inconsistencies between the \texttt{MIST} isochrones and our seismic models, which arise from differences in the adopted input physics. Additionally, a significant discrepancy was found between the \texttt{MIST}-derived and seismic masses for higher-mass \ca{$\gamma\,$Dor pulsators}, with the seismic masses being approximately $0.3\,\mathrm{M_\odot}$ lower than the \ca{mass estimates from \texttt{MIST}.}

To investigate whether the age and mass discrepancies can be mitigated by adopting consistent input physics, we constructed our own isochrones, which are calibrated using \ca{the} asteroseismic results. \ca{This results in a high consistency between isochronal and asteroseismic modelling, where we accounted for the different values of $\eta$ and $f_\mathrm{CBM}$ for the different pulsators.} The newly derived extinction value ($A_0=0.37\pm0.03\,\mathrm{mag}$) \ca{from the new isochronal modelling} is more consistent with independent observational estimates than the \texttt{MIST}-based value, and the newly derived age (\selfisochroneage) shows that the age discrepancy between the \texttt{MIST} 
isochrone and \ca{the joint asteroseismic age of the pulsators} was eliminated. However, the mass discrepancy \ca{was} only partially resolved, indicating a persistent \ca{modest} mismatch between the 
\ca{stellar core and envelope physics in the models and the observed properties of the stars.}

\ca{Our novel methodology and its application to NGC\,2516 opens up additional applications of powerful young open cluster asteroseismology, with Gaia and TESS \cite[or in the future PLATO,][]{Rauer2025} data as observational input. We have shown that our asteroseismic cluster modelling method leads to an age estimation with a superb 6\% relative precision.}

\begin{acknowledgements}
\ca{The authors are grateful for valuable discussion with Dr. Aaron Dotter, particularly for his encouragement to construct asteroseismically based isochrones.
The research leading to these results has received funding from the European Research Council (ERC) under the Horizon Europe programme (Synergy Grant agreement N$^\circ$101071505: 4D-STAR).  While funded by the European Union, views and opinions expressed are however those of the author(s) only and do not necessarily reflect those of the European Union or the European Research Council. Neither the European Union nor the granting authority can be held responsible for them. 
GL acknowledges the Research Foundation Flanders (FWO) for a short stay abroad grant to attend the MESA Down Under School (grant K224824N), and travel support from the National Natural Science Foundation of China (NSFC) through grant 12273002 and the key project 12233013. 
The computational resources and services used in this work were provided by the VSC (Flemish Supercomputer Center), funded by the Research Foundation Flanders (FWO) and the Flemish Government.}
\end{acknowledgements}

%
   \bibliographystyle{aa} 
   \bibliography{aa56409-25} 

\begin{appendix}

\section{Impact of mixing length $\alpha_\mathrm{MLT}$}\label{app_sec:MLT}

\connytwo{\citet{Aerts2018ApJS} studied the impact of changing the input physics for gravito-inertial asteroseismology, including the mixing length parameter. They offered a hierarchy in the importance of various phenomena. They found that changing the mixing length parameter hardly matters for stars with a radiative envelope, but it is of relevance for AF-type pulsators due to their thin convective envelope. However, adopting a meaningful range in values for the mixing length parameter was found to be inferior in importance compared to the inclusion of the Coriolis force. }

\LGtwo{To investigate the actual impact of different values of the mixing length parameter $\alpha_\mathrm{MLT}$ for our application, we recalculated the best-fitting evolutionary tracks for the two stars TIC\,308307454 and TIC\,281582674 using $\alpha_\mathrm{MLT}=2.0$. For the model of TIC\,308307454, we adopted $M=1.54\,\mathrm{M_\odot}$, $f_\mathrm{CBM}=0.005$, and $\eta=1000$, while for TIC\,281582674 we adopted $M=1.78\,\mathrm{M_\odot}$, $f_\mathrm{CBM}=0.030$, and $\eta=1000$. Both models were evolved up to an age of 300\,Myr.}

\LGtwo{Figure~\ref{fig:alpha_HR_TIC308307454} shows the evolutionary tracks of TIC\,308307454 at the beginning of the main sequence. As the star still has a shallow convective envelope, we note that a lower $\alpha_\mathrm{MLT}$ value shifts the evolutionary track towards lower temperatures, consistent with what \cite{Joyce2023} reported using both MESA and the DSEP \citep[Dartmouth Stellar Evolution Program,][]{Dotter2008} code. Such a shift ($\sim6\,\mathrm{K}$) is negligible compared with the observational uncertainties. }

\LGtwo{Figure~\ref{fig:alpha_Pi0_TIC308307454} displays the evolution of the asymptotic spacing $\Pi_0$ as a function of time. We find that a lower $\alpha_\mathrm{MLT}$ value leads to a decrease in $\Pi_0$. This downward shift ($\sim 10\,\mathrm{s}$) is negligible compared with observational uncertainties, confirming that $\alpha_\mathrm{MLT}$ does not strongly affect the convective core.
Figure~\ref{fig:alpha_period_spacing_TIC308307454} compares the period-spacing patterns of the models with $\alpha_\mathrm{MLT}=1.8$ and $2.0$. We find that the two models produce nearly identical period spacings. Therefore, varying $\alpha_\mathrm{MLT}$ does not introduce any significant changes in our modelling.}

\begin{figure}
    \centering
    \includegraphics[width=0.8\linewidth]{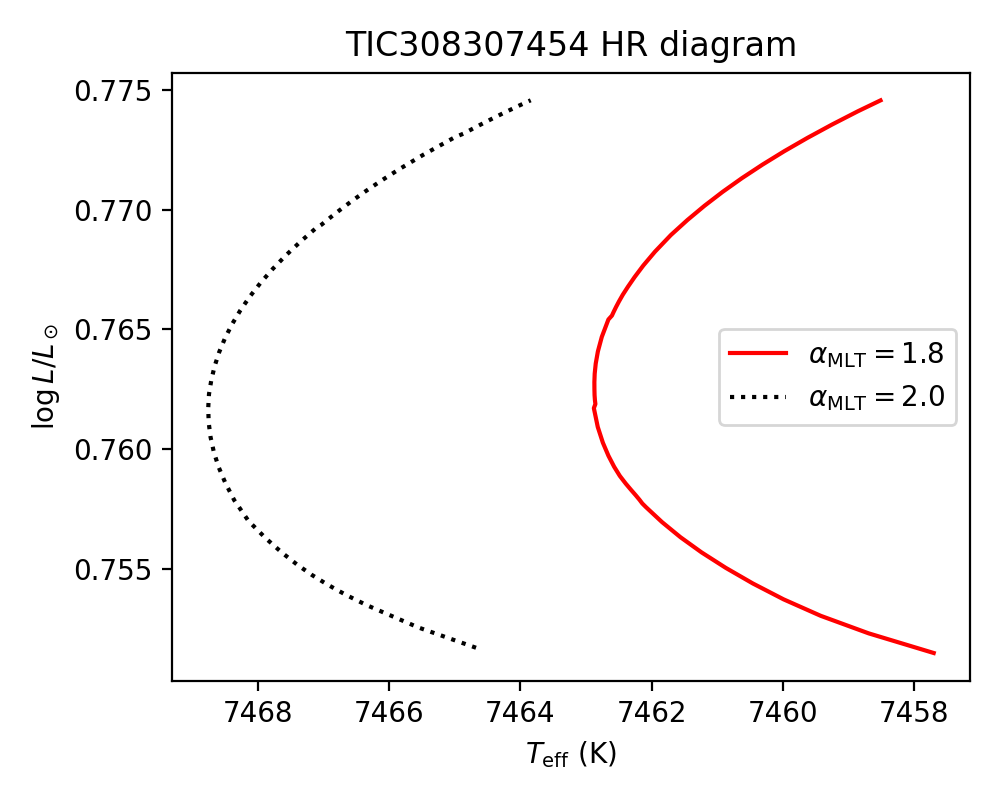}
    \caption{HR diagram showing the evolutionary tracks of the best-fitting model of TIC\,308307454 computed with different values of the mixing length parameter $\alpha_\mathrm{MLT}$. }
    \label{fig:alpha_HR_TIC308307454}
\end{figure}

\begin{figure}
    \centering
    \includegraphics[width=0.8\linewidth]{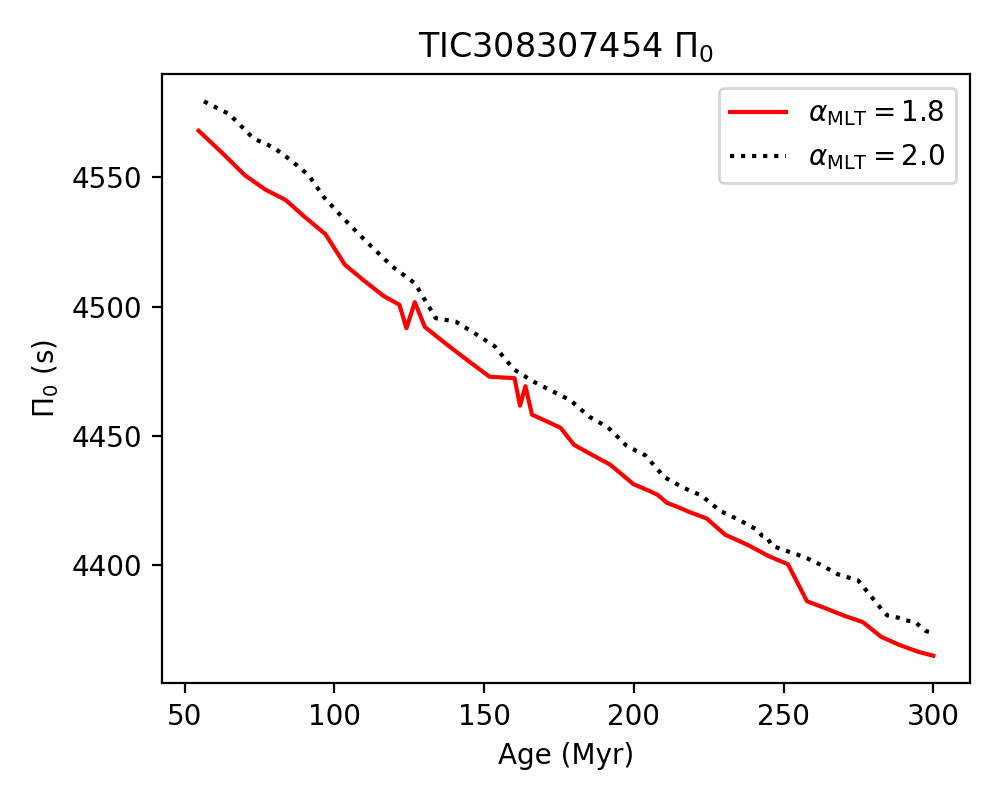}
    \caption{Evolution of the asymptotic period spacing $\Pi_0$ with stellar age for the best-fitting model of TIC\,308307454, using different $\alpha_\mathrm{MLT}$. }
    \label{fig:alpha_Pi0_TIC308307454}
\end{figure}

\begin{figure}
    \centering
    \includegraphics[width=0.8\linewidth]{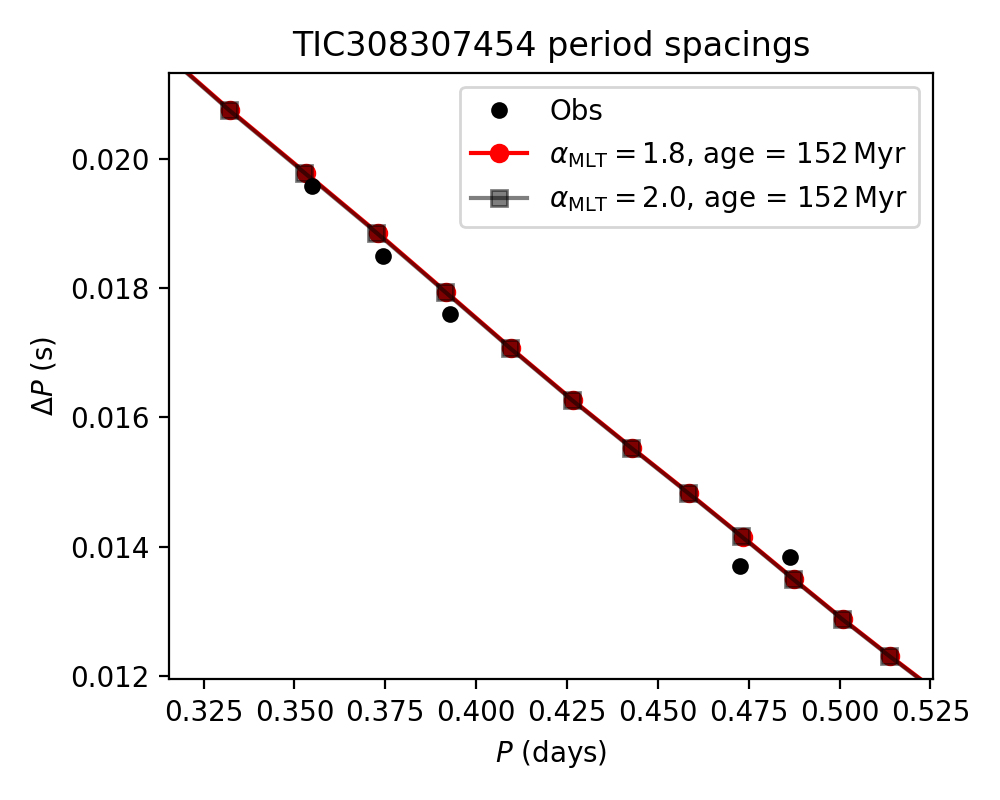}
    \caption{Period-spacing patterns of the best-fitting model of TIC\,308307454 for different values of $\alpha_\mathrm{MLT}$. The g-mode periods are nearly identical, such that the red and gray points overlap.}
    \label{fig:alpha_period_spacing_TIC308307454}
\end{figure}

\LGtwo{As shown in Fig.~\ref{fig:alpha_HR_TIC281582674}, the evolutionary tracks on the HR diagram for TIC\,281582674
exhibit even smaller differences than in Fig.~\ref{fig:alpha_HR_TIC308307454}, because it has a higher mass and a thinner convective envelope. For the evolution of $\Pi_0$ (Fig.~\ref{fig:alpha_Pi0_TIC281582674}), we find that a lower $\alpha_\mathrm{MLT}$ increases $\Pi_0$, in contrast to the trend seen in Fig.~\ref{fig:alpha_Pi0_TIC308307454}.}
\LGtwo{When inspecting the period spacing pattern of TIC\,281582674’s models with different values of $\alpha_\mathrm{MLT}$, we find that adopting $\alpha_\mathrm{MLT}=1.8$ shifts the time at which the dip appears, requiring an older age to reproduce the observed dip. Varying $\alpha_\mathrm{MLT}$ slightly alters the internal structure and such structural changes may impact the modes in a variety of ways, depending on the nature of the core (growing or shrinking) and the thickness of the convective envelope. Some modes are hardly affected while others are quite sensitive to changes in the mode cavity. This notably matters for trapped modes, as they are dependent on the structure in the boundary layer between the convective core and the radiative envelope \citep[e.g.][]{Miglio2008MNRAS,Michielsen2021,Vanlaer2023}. 
We have not investigated the relation or degeneracy between $\alpha_\mathrm{MLT}$ and the other parameters because TIC\,281582674 reveals only one tiny dip. }


\begin{figure}
    \centering
    \includegraphics[width=0.8\linewidth]{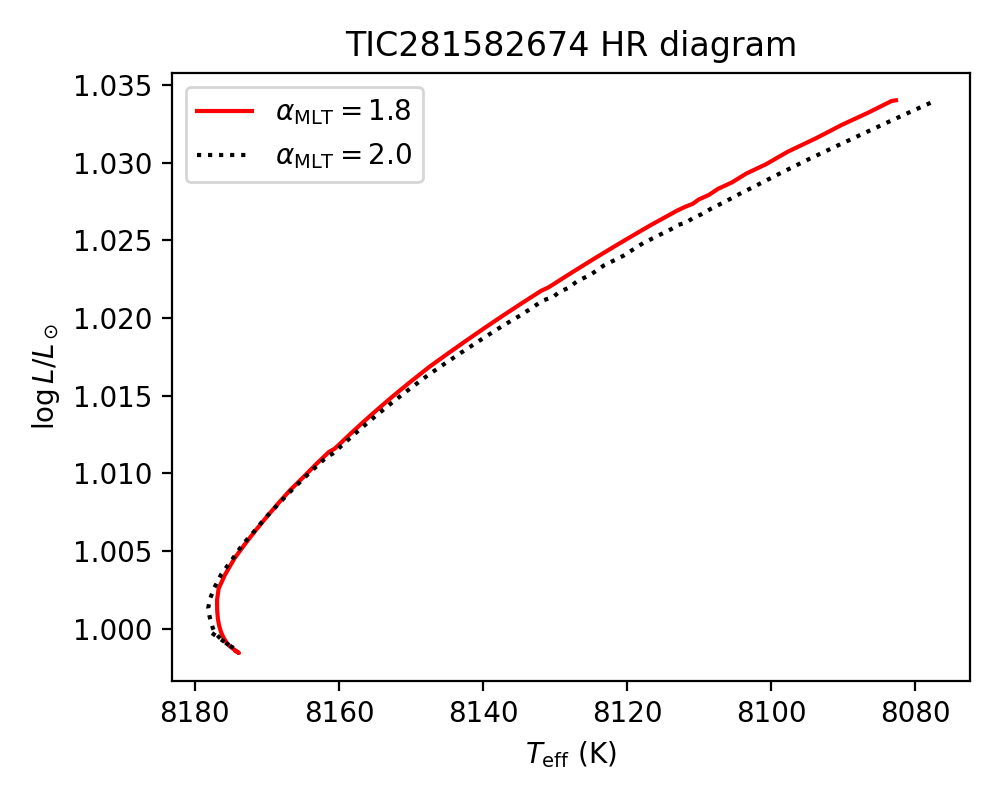}
    \caption{HR diagram showing the evolutionary tracks of the best-fitting model of TIC\,281582674 computed with different values of the mixing length parameter $\alpha_\mathrm{MLT}$. }
    \label{fig:alpha_HR_TIC281582674}
\end{figure}

\begin{figure}
    \centering
    \includegraphics[width=0.8\linewidth]{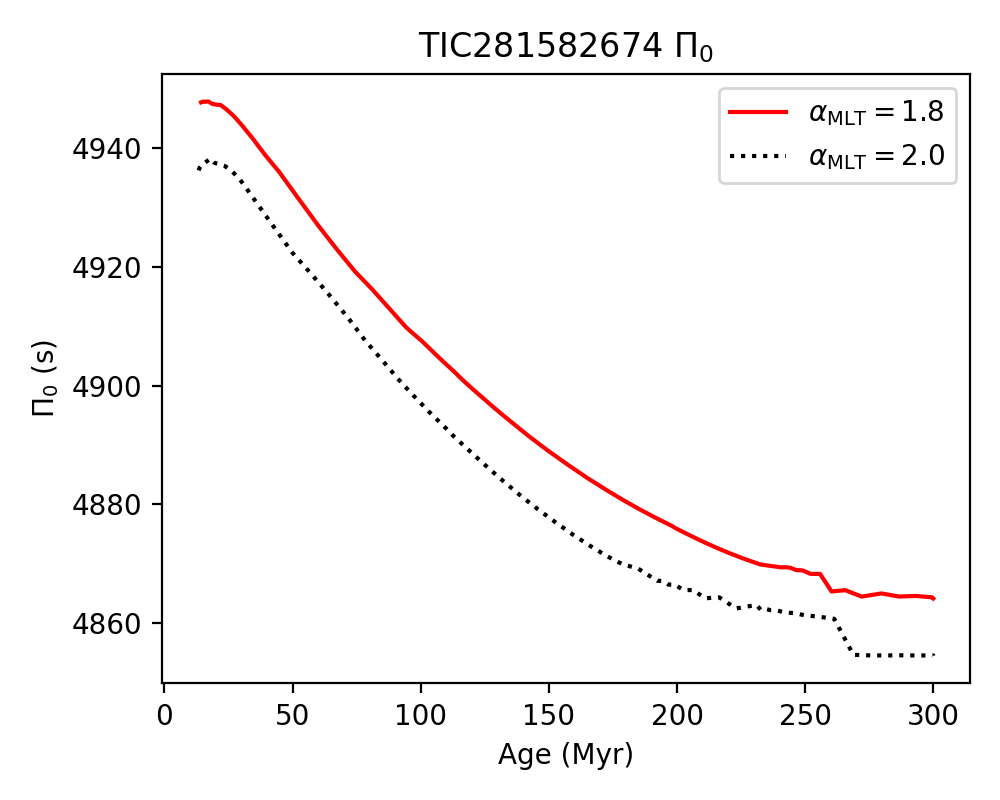}
    \caption{Evolution of the asymptotic period spacing $\Pi_0$ with stellar age for the best-fitting model of TIC\,281582674, using different $\alpha_\mathrm{MLT}$. }
    \label{fig:alpha_Pi0_TIC281582674}
\end{figure}

\begin{figure}
    \centering
    \includegraphics[width=0.8\linewidth]{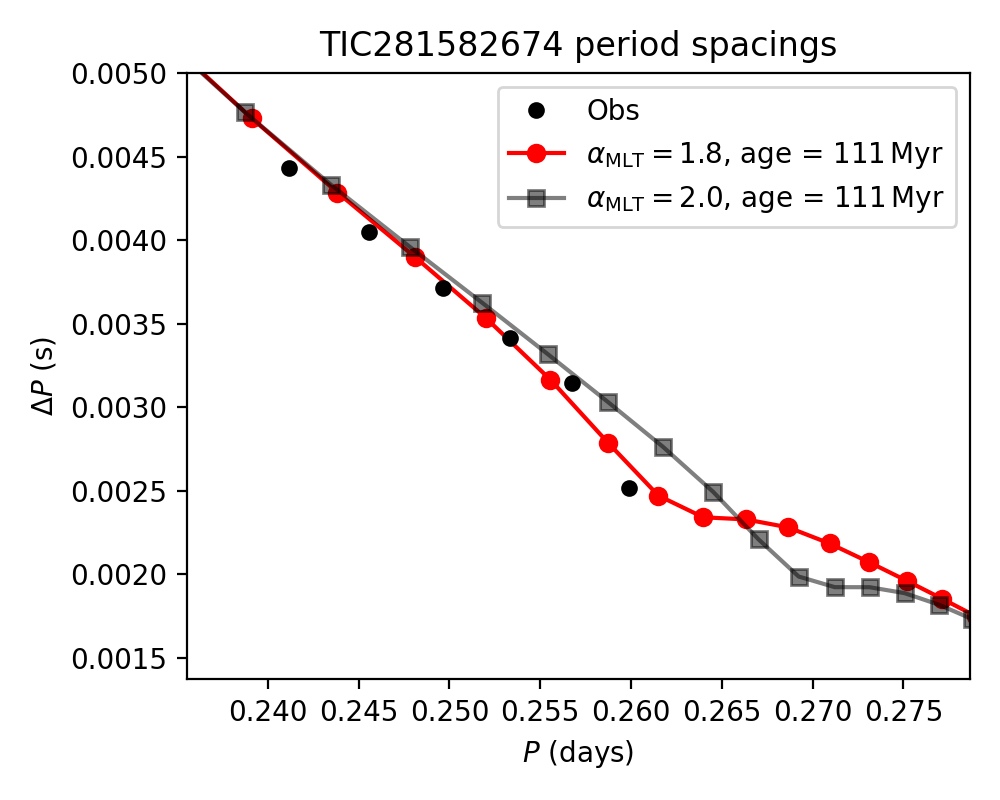}
    \caption{Period-spacing patterns of the best-fitting model of TIC\,281582674 for different values of $\alpha_\mathrm{MLT}$. 
}
    \label{fig:alpha_period_spacing_TIC281582674}
\end{figure}

\section{Fitting results for TIC\,358466729 and TIC\,410452218 using Approach~1}\label{app_sec:fitting_approach1}

Here we present the posterior distributions and best-fitting period spacing patterns of TIC\,358466729 and TIC\,410452218 obtained using Approach~1. These two stars have limited ability to constrain age and other stellar parameters, as $\Pi_0$ evolves slowly with time in their mass ranges and no clear dips are observed in their period spacing patterns. TIC\,358466729 exhibits Rossby modes with $k = -2$ and $m = -1$ \citep{Saio2018_r_modes, Li2019_r_mode}.

\begin{figure}
    \centering
    \includegraphics[width=0.9\linewidth]{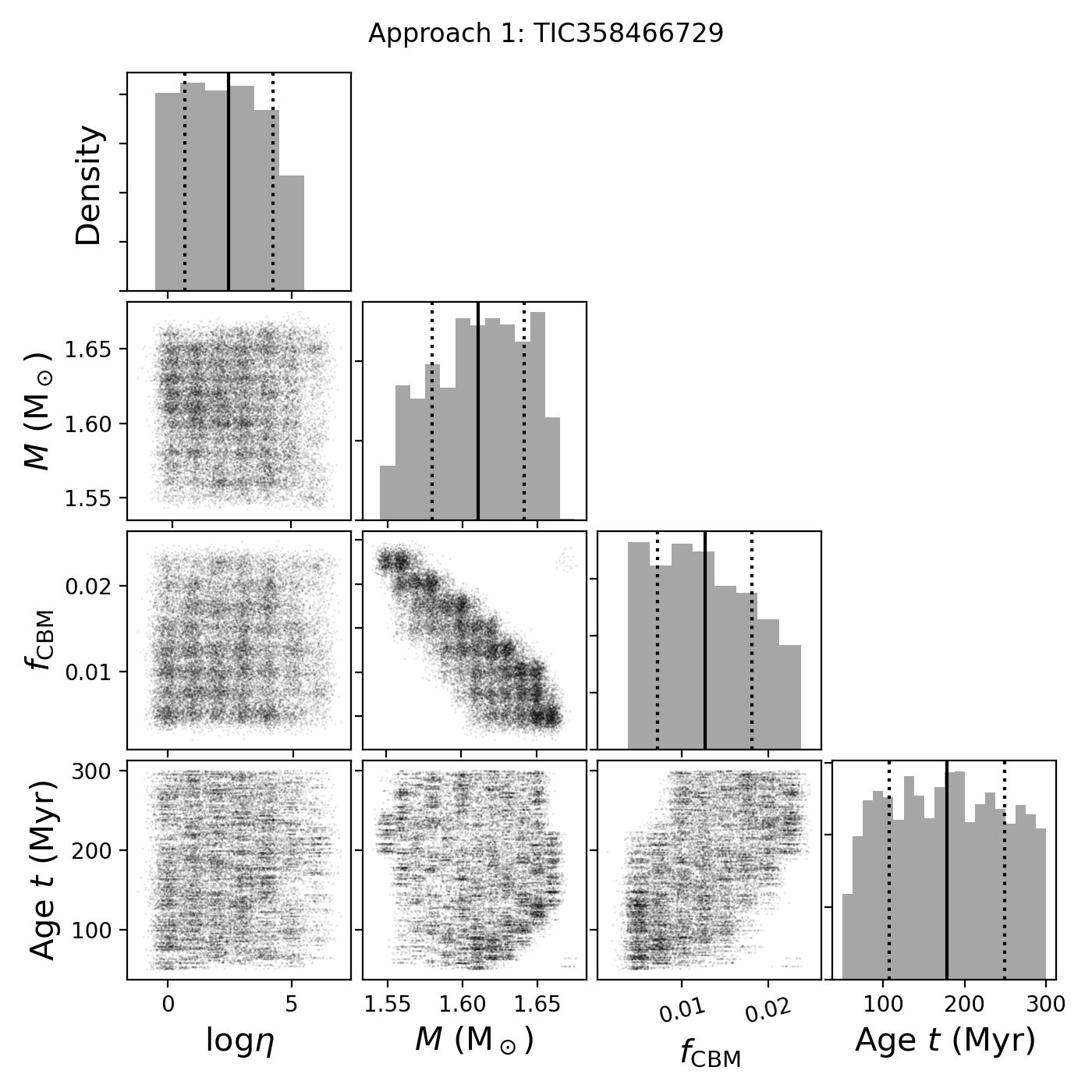}
    \caption{Same as Fig.~\ref{fig:corner_TIC308307454}, but for TIC\,358466729. }
    \label{appen_appro1_fig:corner_TIC358466729}
\end{figure}

\begin{figure}
    \centering
    \includegraphics[width=0.9\linewidth]{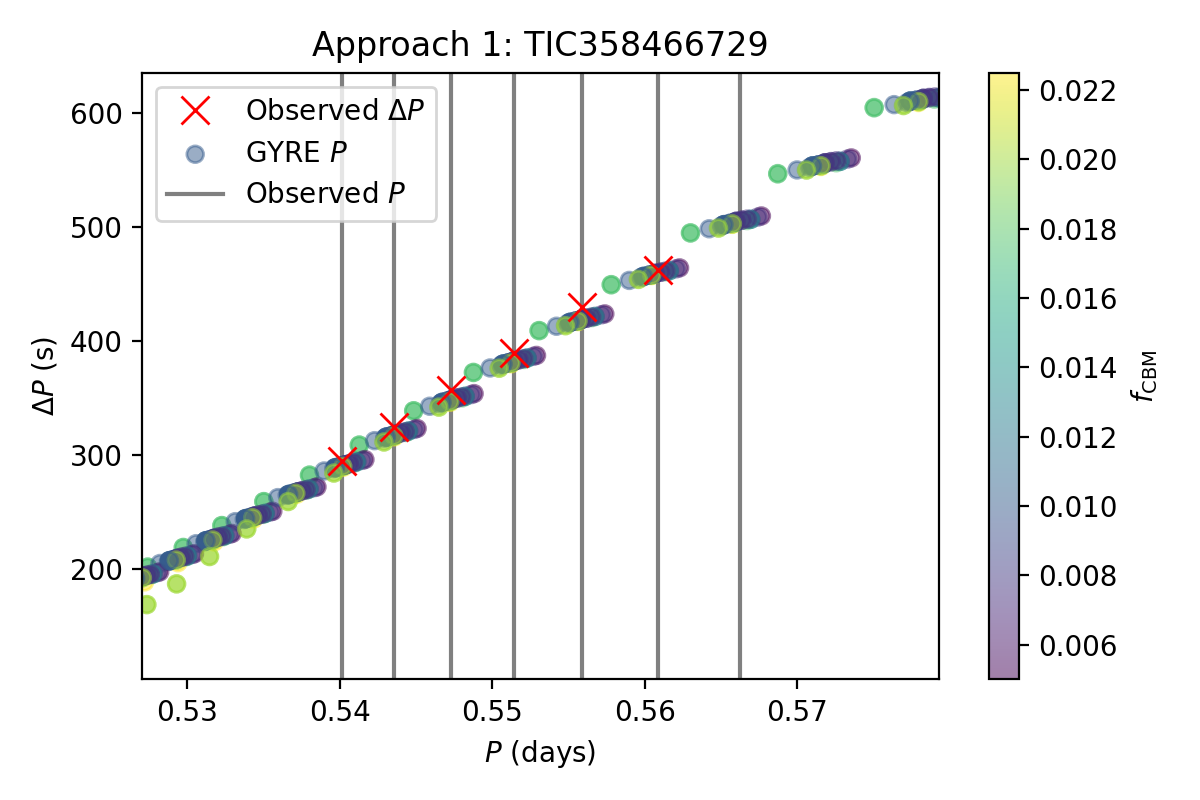}
    \caption{Same as Fig.~\ref{fig:DP_approach1_TIC308307454}, but for TIC\,358466729. This star exhibit $k=-2,\,m=-1$ r modes. }
    \label{appen_appro1_fig:DP_approach1_TIC358466729}
\end{figure}

\begin{figure}
    \centering
    \includegraphics[width=0.9\linewidth]{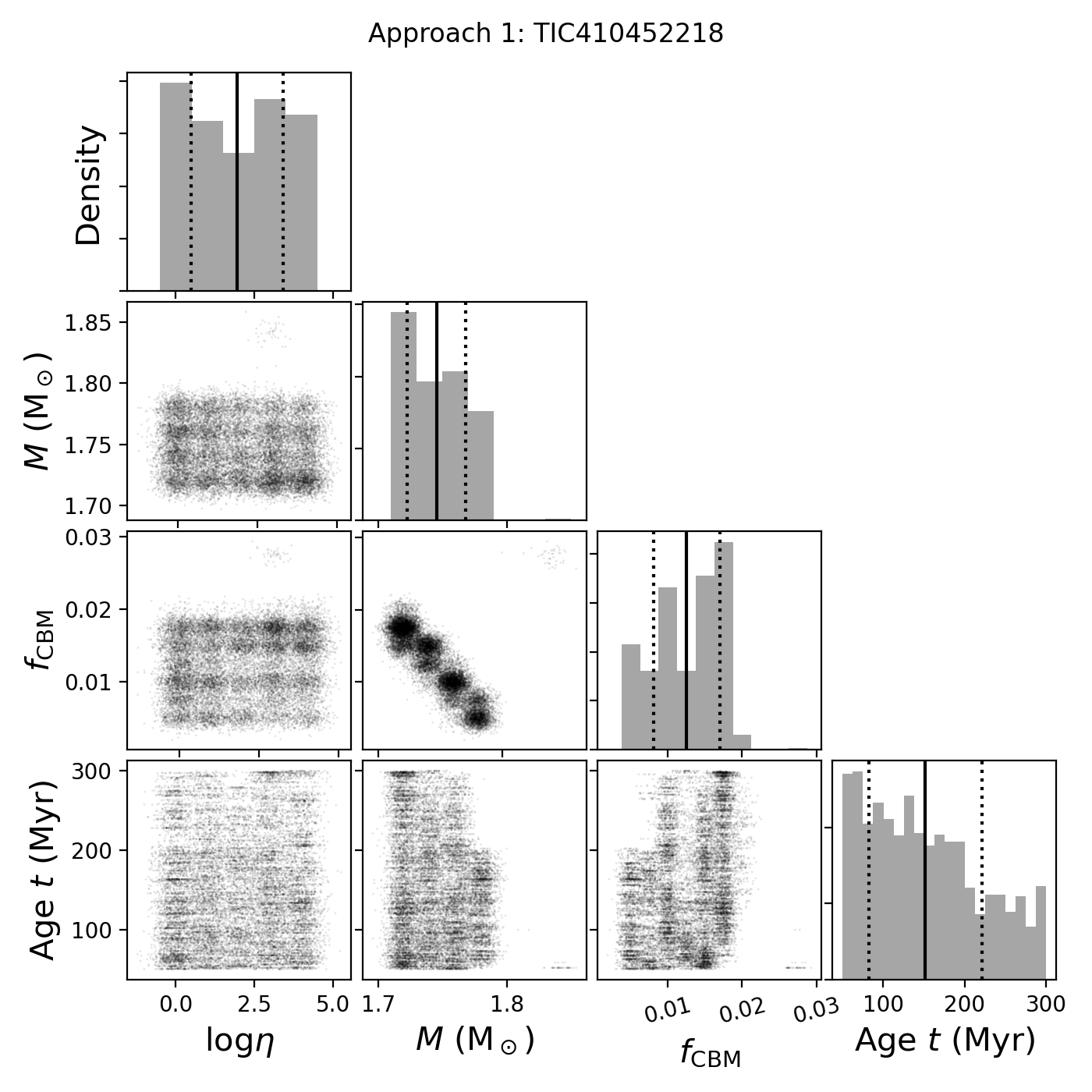}
    \caption{Same as Fig.~\ref{fig:corner_TIC308307454}, but for TIC\,410452218. }
    \label{appen_appro1_fig:corner_TIC410452218}
\end{figure}

\begin{figure}
    \centering
    \includegraphics[width=0.9\linewidth]{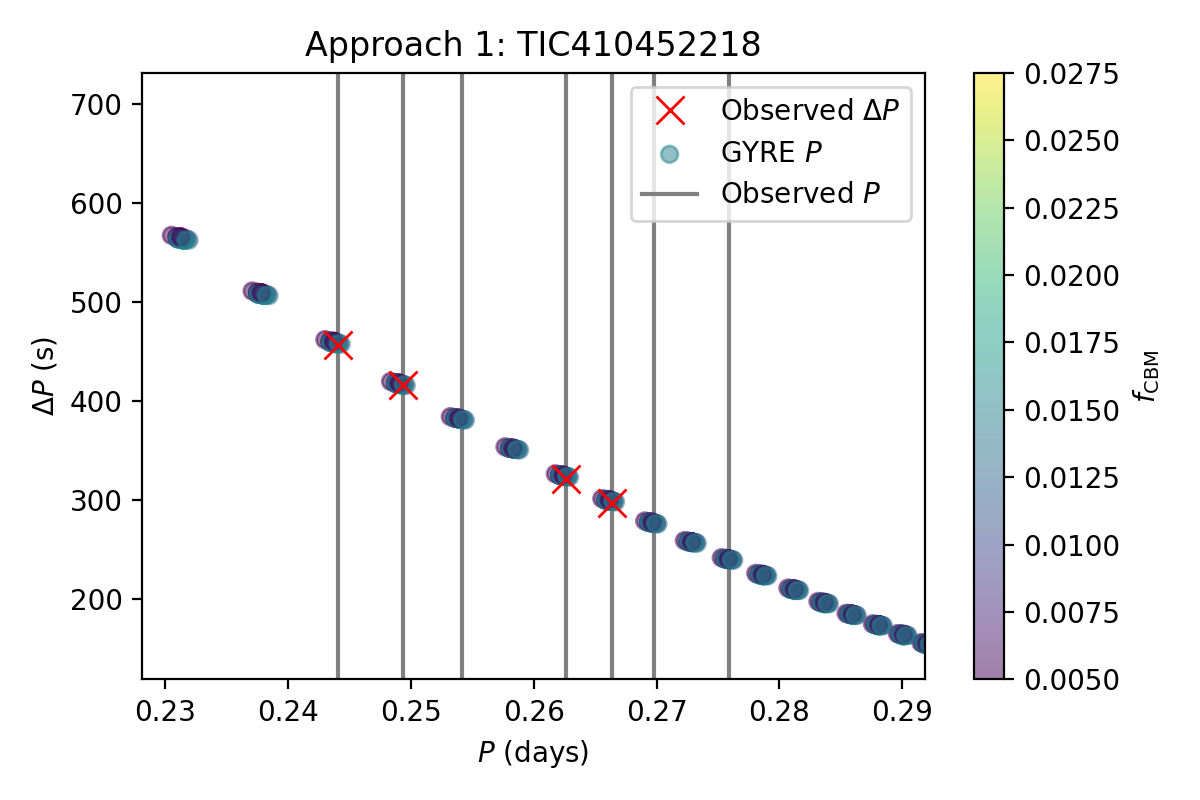}
    \caption{Same as Fig.~\ref{fig:DP_approach1_TIC308307454}, but for TIC\,410452218. }
    \label{appen_appro1_fig:DP_approach1_TIC410452218}
\end{figure}

\section{Fitting results using Approach~2}\label{app_sec:fitting_approach2}

In contrast to Approach~1 shown above, Approach~2 imposes that all four stars share the same age, which is more consistent with the condition of open clusters. We present the posterior distributions and the best-fitting period spacing patterns here.

\begin{figure}
    \centering
    \includegraphics[width=0.9\linewidth]{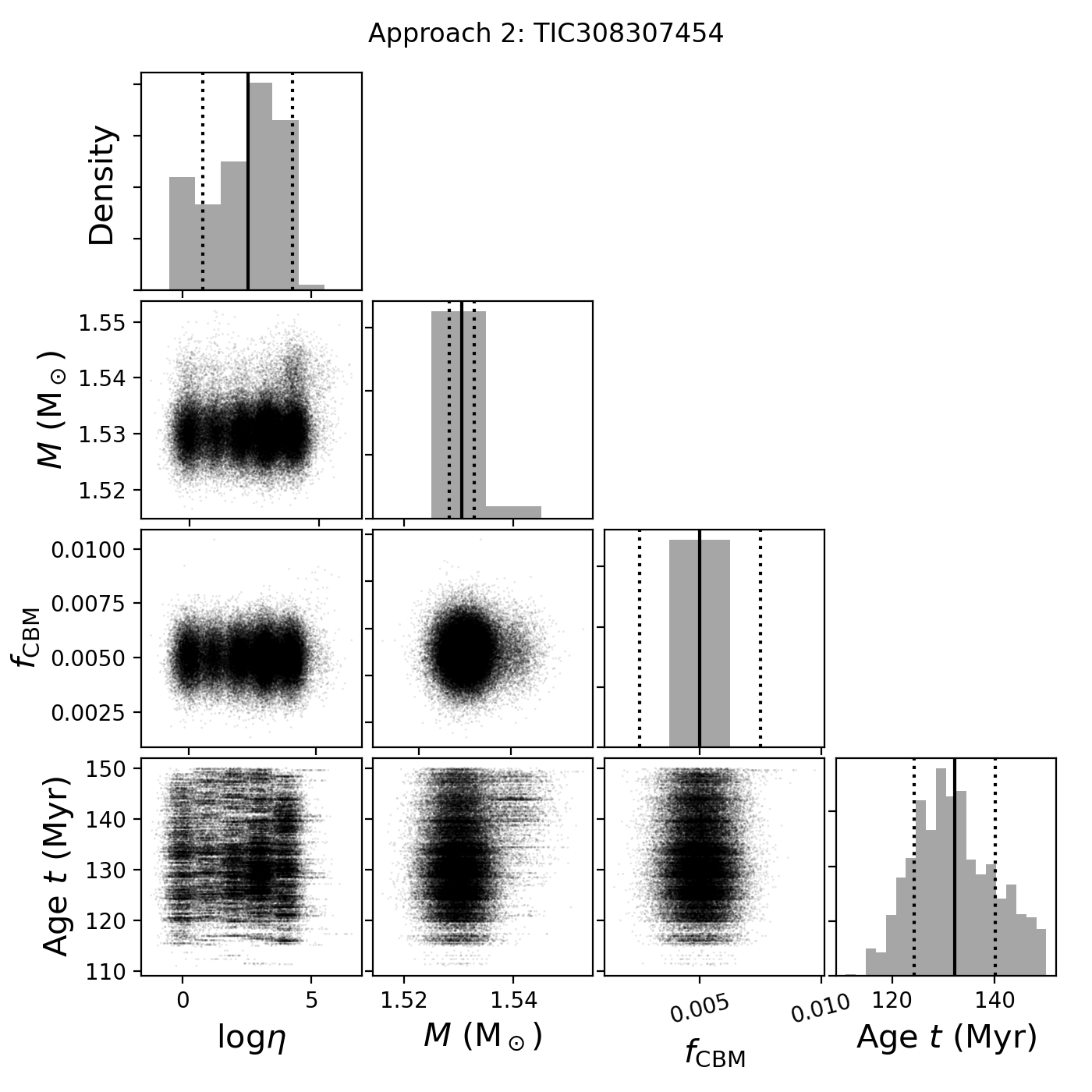}
    \caption{Same as Fig.~\ref{fig:corner_TIC308307454} for TIC\,308307454, but using Approach~2. The bottom-right panel shows the distributions of the joint age derived using Approach~2, so this panel is identical in Figs.~\ref{appen_appro2_fig:corner_TIC358466729}, \ref{appen_appro2_fig:corner_TIC410452218}, and \ref{appen_appro2_fig:corner_TIC281582674}. }
    \label{appen_appro2_fig:corner_TIC308307454}
\end{figure}

\begin{figure}
    \centering
    \includegraphics[width=0.9\linewidth]{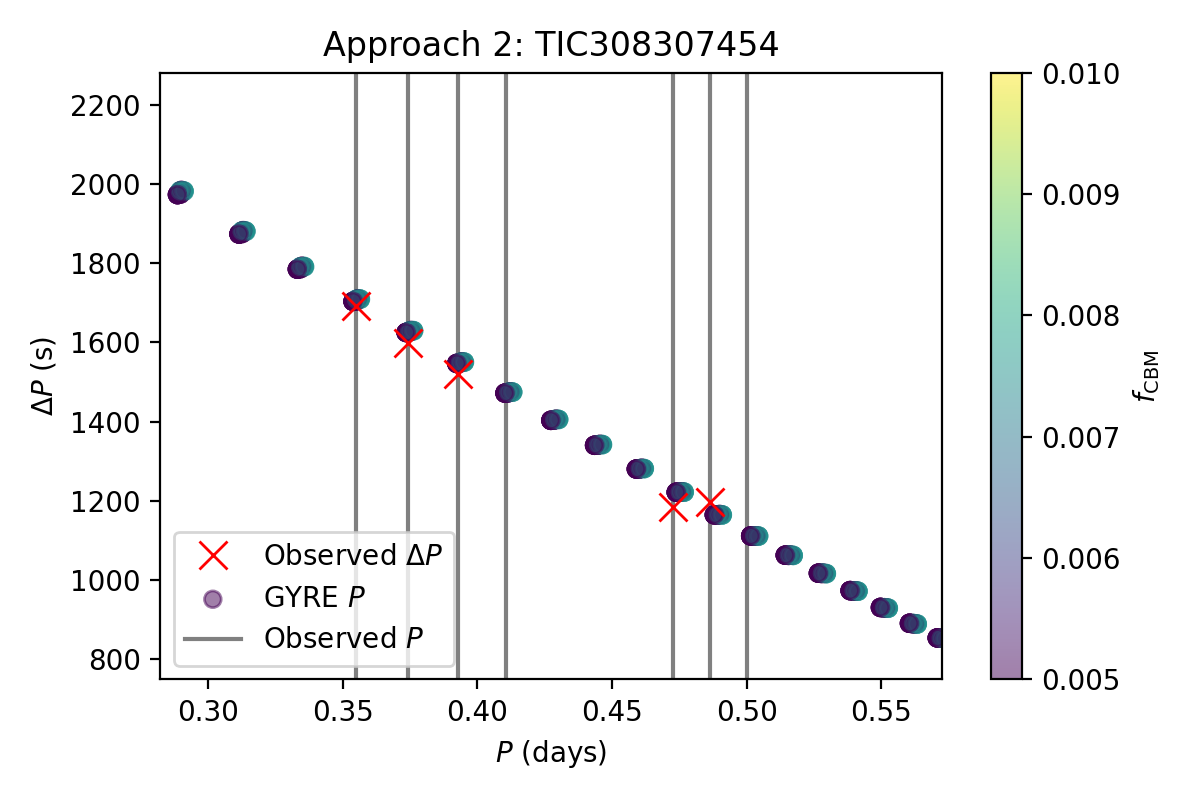}
    \caption{Same as Fig.~\ref{fig:DP_approach1_TIC308307454} for TIC\,308307454, but using Approach~2.}
    \label{appen_appro2_fig:DP_approach2_TIC308307454}
\end{figure}

\begin{figure}
    \centering
    \includegraphics[width=0.9\linewidth]{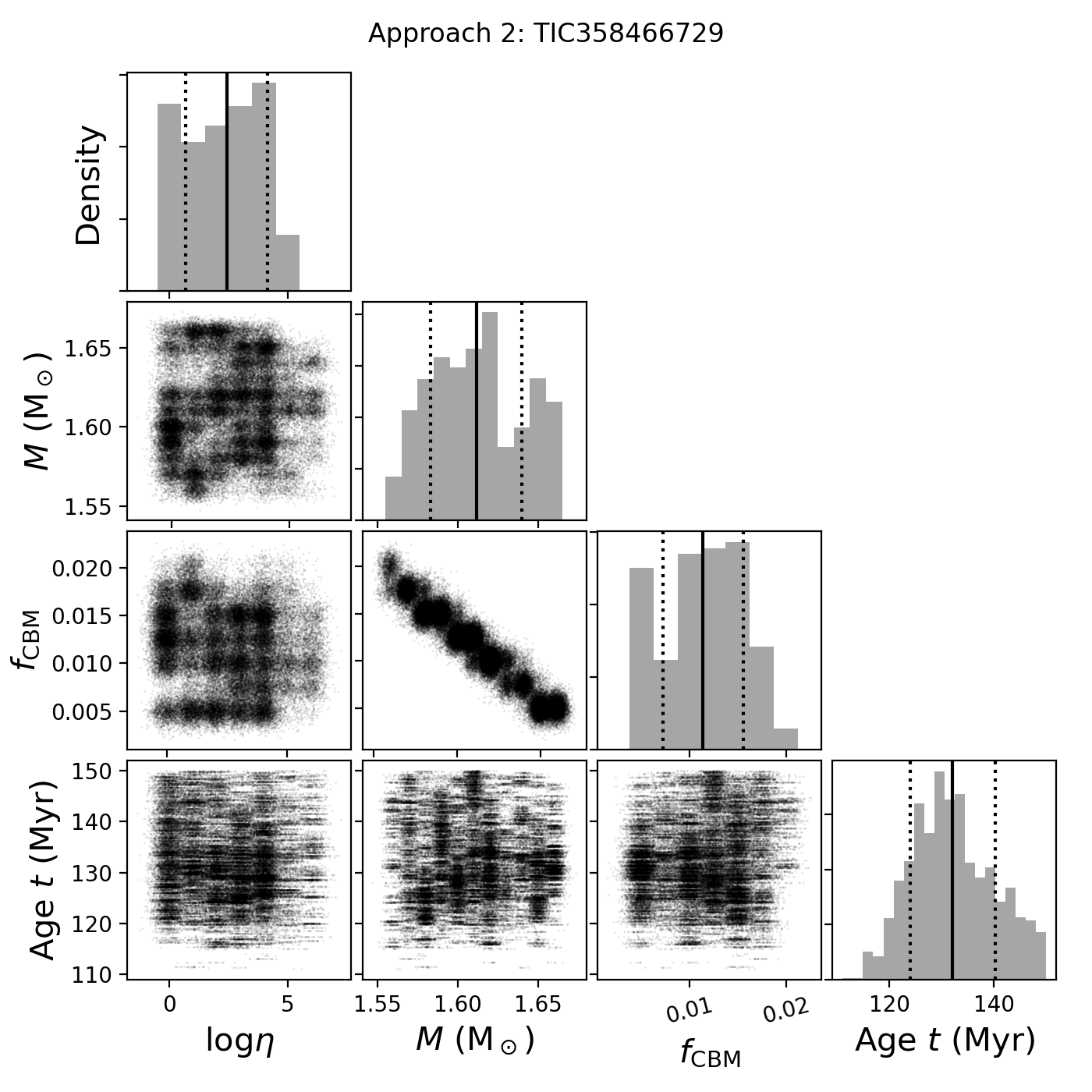}
    \caption{Same as Fig.~\ref{fig:DP_approach1_TIC308307454}, but for TIC\,358466729 using Approach~2.}
    \label{appen_appro2_fig:corner_TIC358466729}
\end{figure}

\begin{figure}
    \centering
    \includegraphics[width=0.9\linewidth]{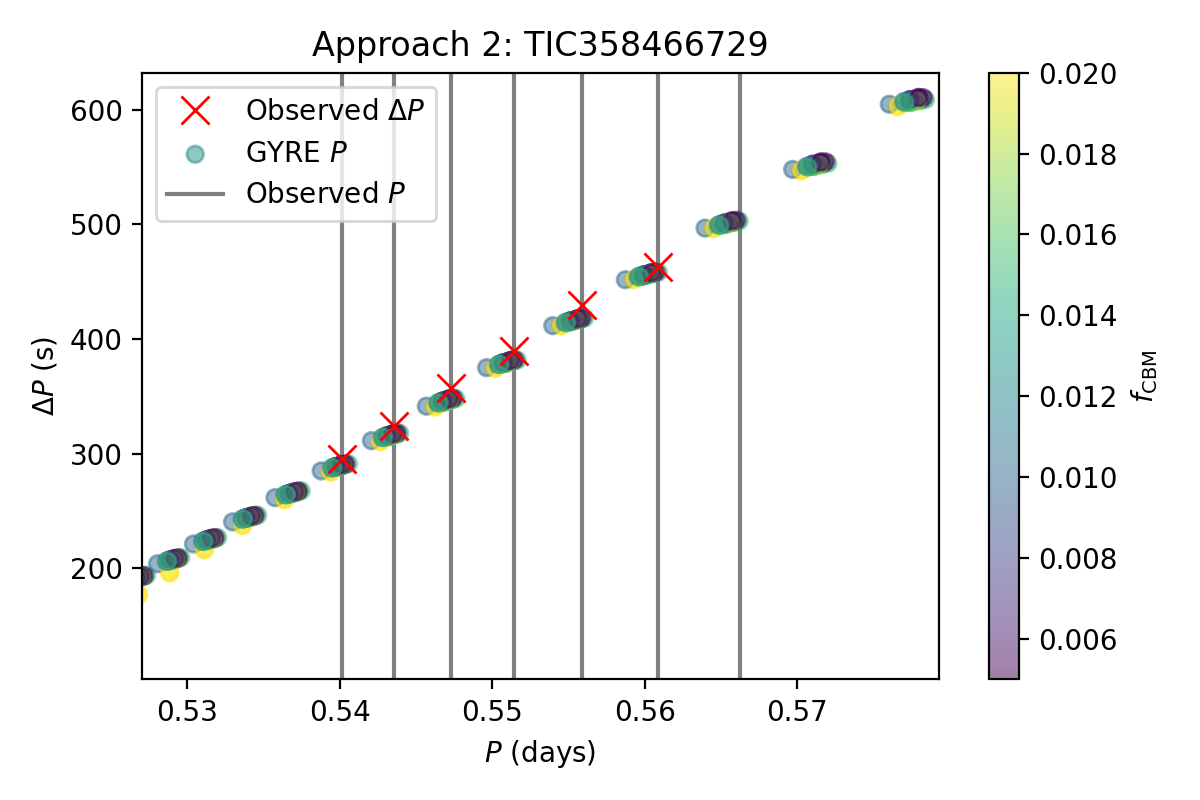}
    \caption{Same as Fig.~\ref{fig:DP_approach1_TIC308307454}, but for TIC\,358466729 using Approach~2.}
    \label{appen_appro2_fig:DP_approach2_TIC358466729}
\end{figure}

\begin{figure}
    \centering
    \includegraphics[width=0.9\linewidth]{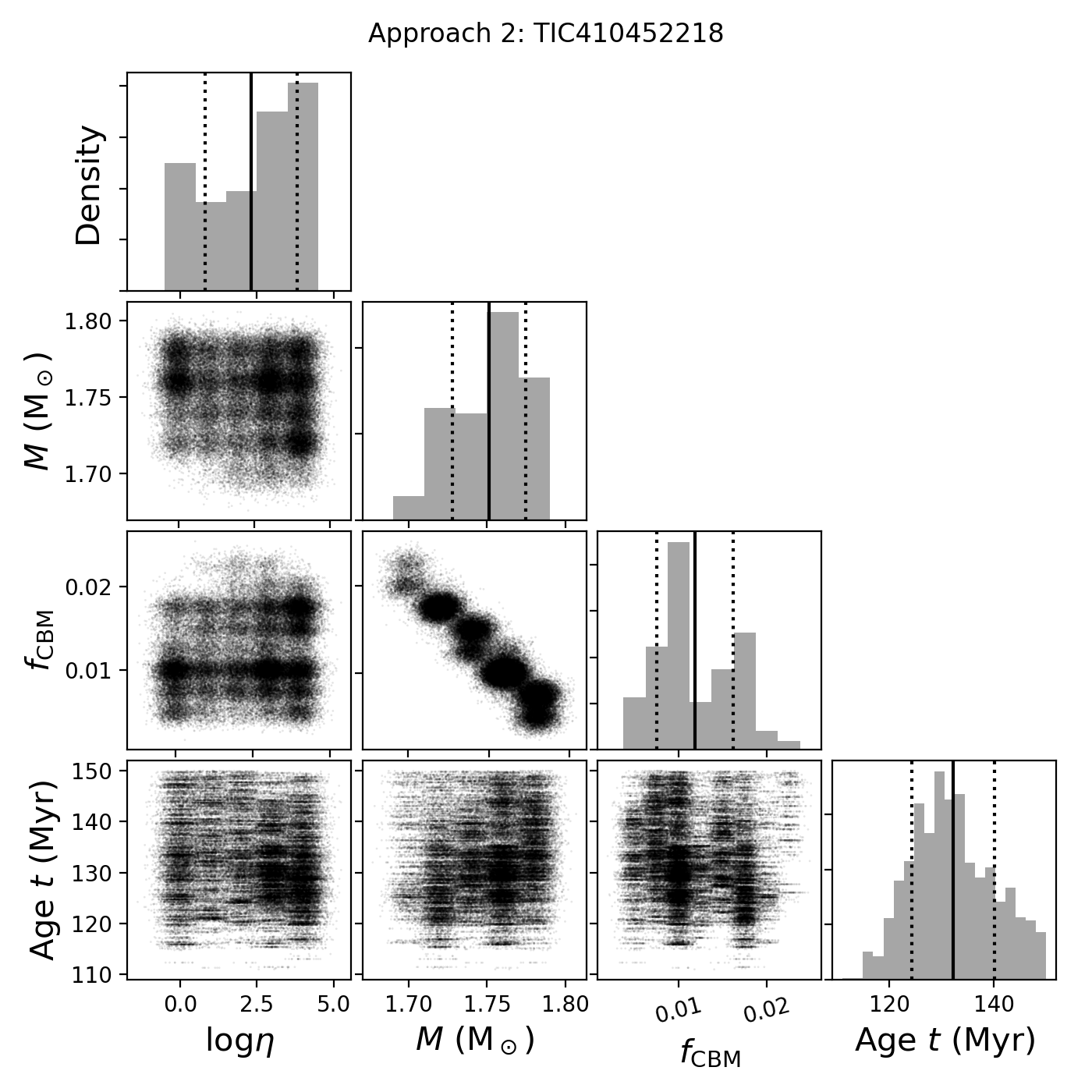}
    \caption{Same as Fig.~\ref{fig:DP_approach1_TIC308307454}, but for TIC\,410452218 using Approach~2.}
    \label{appen_appro2_fig:corner_TIC410452218}
\end{figure}

\begin{figure}
    \centering
    \includegraphics[width=0.9\linewidth]{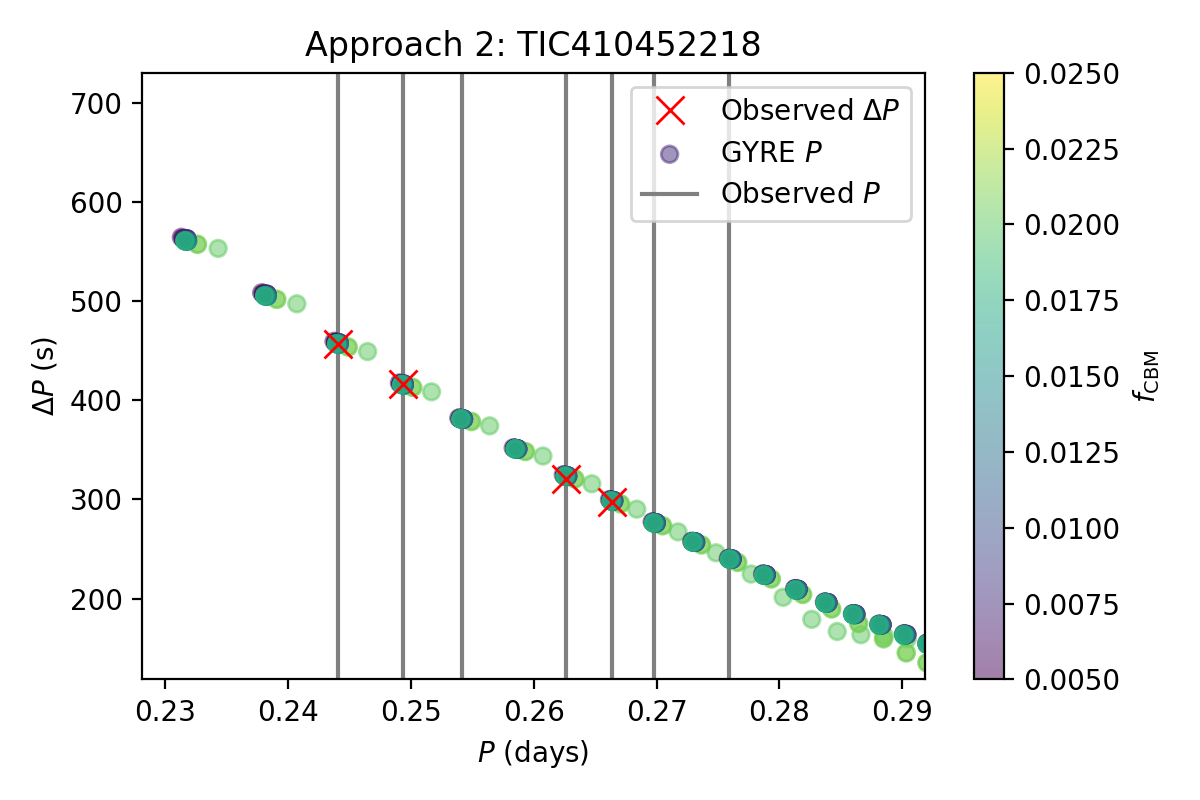}
    \caption{Same as Fig.~\ref{fig:DP_approach1_TIC308307454}, but for TIC\,410452218 using Approach~2.}
    \label{appen_appro2_fig:DP_approach2_TIC410452218}
\end{figure}

\begin{figure}
    \centering
    \includegraphics[width=0.9\linewidth]{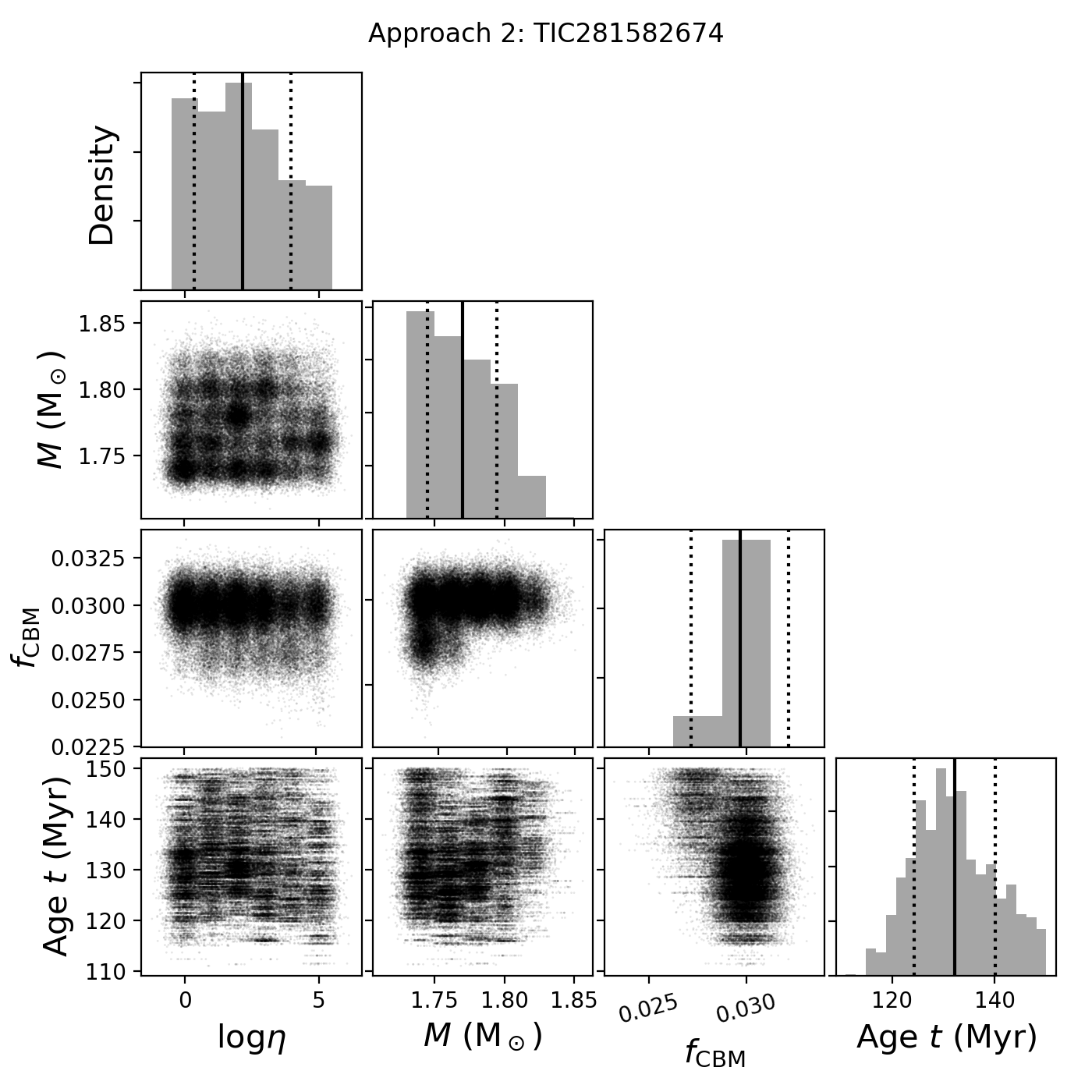}
    \caption{Same as Fig.~\ref{fig:DP_approach1_TIC308307454}, but for TIC\,281582674 using Approach~2.}
    \label{appen_appro2_fig:corner_TIC281582674}
\end{figure}

\begin{figure}
    \centering
    \includegraphics[width=0.9\linewidth]{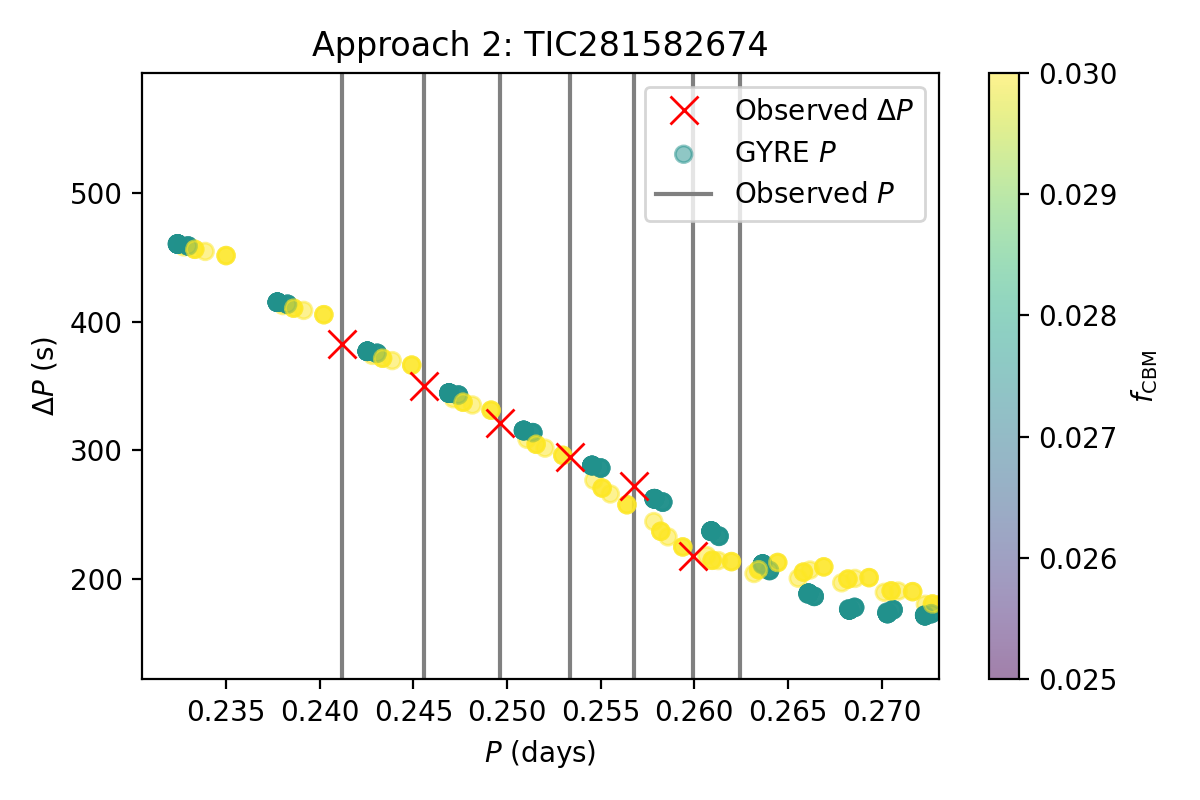}
    \caption{Same as Fig.~\ref{fig:DP_approach1_TIC308307454}, but for TIC\,281582674 using Approach~2.}
    \label{appen_appro2_fig:DP_approach2_TIC281582674}
\end{figure}

\end{appendix}

\end{document}